\documentclass{llncs} 

\pdfoutput = 1
\usepackage{verbatim}
\usepackage{amsmath}
\usepackage{graphicx}
\usepackage{caption}
\usepackage{algpseudocode}
\usepackage{color,xparse}
\usepackage{wasysym}
\usepackage{array}
\usepackage{rotating}
\usepackage{setspace}
\usepackage{longtable}
\usepackage{rotating}
\usepackage{subfig}
\usepackage{epsfig}
\usepackage{epstopdf}
\usepackage{ragged2e}
\newcolumntype{Y}{>{\RaggedRight\arraybackslash}X} 
\usepackage{tabularx}
\usepackage{amsfonts}
\usepackage{amssymb}
\usepackage{cite}
\usepackage{wrapfig}
\usepackage{upgreek}
\usepackage{dsfont}
\usepackage[toc,page]{appendix}
\usepackage{latexsym}
\usepackage{fancybox}
\usepackage{amsfonts} 
\usepackage[T1]{fontenc} 
\DeclareGraphicsExtensions{.pdf,.eps,.jpg,.png}
\usepackage{listings} 
\usepackage{textcomp}
\usepackage[font={small,it}]{caption}
\usepackage{url}
\usepackage{enumitem}

\newcounter{MyCounter}

\title{Improved Spanning Ratio for Low Degree Plane Spanners\thanks{This work was partially supported by the Natural Sciences and Engineering Research Council of Cananda (NSERC) and by the Ontario Graduate Scholarship (OGS).} }
\author{Prosenjit Bose \and Darryl Hill \and Michiel Smid}

\institute{School of Computer Science, Carleton University. Ottawa, Canada}
\begin{document}

\newcommand{\ov}[1]{\overline{#1}}
\newcommand{\halfspan}{|pa|+\frac{\theta}{\sin\theta}|aq|}
\newcommand{\func}{\frac{\theta}{\sin\theta}}
\newcommand{\fullspan}[1]{\left(1+\frac{\theta}{\sin\theta}
	\right)#1}
\newcommand{\idealspan}[1]{\left(1+\frac{2\pi}{3\sqrt{3}}
	\right)#1}
\newcommand{\ospan}{(1+\frac{2\pi}{3\sqrt{3}})}
\newcommand{\tspan}{(1+\frac{\theta}{\sin\theta})}

\newcommand{\nb}{neighbourhood }
\newcommand{\nbsp}{neighbourhoods}
\newcommand{\nbs}{neighbourhoods }
\newcommand{\nbp}{neighbourhood}
\newcommand{\Nb}{Neighbourhood }

\newcommand{\rn}{restricted neighbourhood }
\newcommand{\rnsp}{restricted neighbourhoods}
\newcommand{\rns}{restricted neighbourhoods }
\newcommand{\rnp}{restricted neighbourhood}
\newcommand{\Rn}{Restricted neighbourhood }

\newcommand{\nc}{cone neighbourhood }
\newcommand{\ncsp}{cone neighbourhoods}
\newcommand{\ncs}{cone neighbourhoods }
\newcommand{\ncp}{cone neighbourhood}
\newcommand{\Nc}{Cone neighbourhood }
\newcommand{\Ncs}{Cone neighbourhoods }
\newcommand{\Ncsp}{Cone neighbourhoods}

\newcommand{\rcan}{region }
\newcommand{\rcans}{regions }
\newcommand{\rcansp}{regions}
\newcommand{\rcanp}{region}
\newcommand{\rCan}{Region }
\newcommand{\rCans}{Regions }
\newcommand{\rCaN}{Region }
\newcommand{\rCaNs}{Regions }

\newcommand{\cre}{canonical edge }
\newcommand{\cres}{canonical edges }
\newcommand{\cresp}{canonical edges}
\newcommand{\crep}{canonical edges}
\newcommand{\Cre}{Canonical edge }
\newcommand{\Cres}{Canonical edges }
\newcommand{\CrE}{Canonical Edge }
\newcommand{\CrEs}{Canonical Edges }

\newcommand{\can}{canonical subgraph }
\newcommand{\cans}{canonical subgraphs }
\newcommand{\cansp}{canonical subgraphs}
\newcommand{\canp}{canonical subgraph}
\newcommand{\Can}{Canonical subgraph }
\newcommand{\Cans}{Canonical subgraphs }
\newcommand{\CaN}{Canonical Subgraph }
\newcommand{\CaNs}{Canonical Subgraphs }

\newcommand{\ip}{ideal path }
\newcommand{\ips}{ideal paths }
\newcommand{\ipsp}{ideal paths}
\newcommand{\ipp}{ideal path}
\newcommand{\Ip}{Ideal path }
\newcommand{\Ips}{Ideal paths }
\newcommand{\IP}{Ideal Path }
\newcommand{\IPs}{Ideal Paths }

\newcommand{\EA}{E_A}
\newcommand{\EB}{E_{CAN}}
\newcommand{\EC}{Ec}
\newcommand{\N}[3]{N_{#1}^{#2,#3}}
\newcommand{\pvt}{pivotal }
\newcommand{\C}{Can}

\maketitle	
\begin{abstract}
	We describe an algorithm that builds a plane spanner with a maximum degree of 8 and a spanning ratio of $\approx 4.414$ with respect to the complete graph. This is the best currently known spanning ratio for a plane spanner with a maximum degree of less than 14. 
\end{abstract}

\section{Introduction}

Let $P$ be a set of $n$ points in the plane. Let $G$ be a weighted geometric graph on vertex set $P$, where edges are straight line segments and are weighted according the the Euclidean distance between their endpoints. Let $\delta_{G}(p,q)$ be the sum of the weights of the edges on the shortest path from $p$ to $q$ in $G$. If, for graphs $G$ and $H$ on the point set $P$, where $G$ is a subgraph of $H$, for every pair of points $p$ and $q$ in $P$, $\delta_G(p,q) \leq t\cdot\delta_H(p,q)$ for some real number $t>1$, then $G$ is a $t$-spanner of $H$, and $t$ is called the \emph{spanning ratio}. $H$ is called the \emph{underlying} graph of $G$. In this paper the underlying graph is the Delaunay triangulation or the complete graph. 

The $L_1$-Delaunay triangulation was first proven to be a $\sqrt{10}$-spanner by Chew\cite{chew}. Dobkin \emph{at al.}\cite{dobkin} proved that the $L_2$-Delaunay triangulation is a $\frac{1+\sqrt{5}}{2}\pi$-spanner. This was improved by Keil and Gutwin\cite{keil} to $\frac{2\pi}{3\cos(\frac{\pi}{6})}$, and finally taken to its currently best known spanning ratio of $1.998$ by Xia\cite{xia}.

The Delaunay triangulation may have an unbounded degree. High degree nodes can be detrimental to real world applications of graphs. Thus there has been research into bounded degree plane spanners. We present a brief overview of some of the results in Table \ref{table-uno}. 

\begin{table}
	\vspace{-1cm}
	\begin{center}
	\begin{tabular}{|c|c|c|}
		\hline
		Paper & Degree & Stretch Factor \\
		\hline
		\hline
		Bose \emph{et al.}\cite{bose27} & 27 & $(\pi+1)C_{DT} \approx 8.27$\\
		Li \& Wang\cite{wang23} & 23 & $(1+\pi\sin(\frac{\pi}{4}))C_{DT} \approx 6.44$ \\
		Bose \emph{et al.}\cite{bose17} & 17 & $(\frac{1+\sqrt{3}+3\pi}{2}+2\pi\sin(\pi/12))C_{DT} \approx 23.58$ \\
		Kanj \emph{et al.}\cite{kanj} & 14 & $(1 + \frac{2\pi}{14\cos(\pi/14)})C_{DT} \approx 2.92$ \\
		Bose \emph{et al.}\cite{bose-paz} & 7 & $(\frac{1}{1-2\tan(\pi/8)})C_{DT} \approx  11.65$ \\
		Bose \emph{et al.}\cite{bose-paz} & 6 & $(\frac{1}{(1-\tan(\pi/7)(1+1/\cos(\pi/14)))}C_{DT} \approx 81.66$ \\
		Bonichon \emph{et al.}\cite{bonichon} & 6 & 6 \\
		Bonichon \emph{et al.}\cite{degree4} & 4 & $\sqrt{4+2\sqrt{2}}(19+29\sqrt{2})\approx 156.82$ \\
		\hline \hline
		This paper & 8 & $(1+\frac{2\pi}{6\cos(\pi/6)})C_{DT} \approx 4.41$\\
		\hline
	\end{tabular}\\
	$C_{DT}$ is the spanning ratio of the Delaunay triangulation, currently $< 1.998$\cite{xia}
		\caption{Known results for bounded degree plane spanners.}\label{table-uno}
	\end{center}
	\vspace{-1cm}
\end{table}
Bounded degree plane spanners are often obtained by taking a subset of edges of an existing plane spanner and ensuring that it has bounded degree, while maintaining spanning properties. We note how in Table \ref{table-uno} that all of the results are subgraphs of some variant of the Delaunay triangulation. 

Our contribution is an algorithm to construct a plane spanner of maximum degree 8 with a spanning ratio of $\approx 4.41$. This is the lowest spanning ratio of any graph of degree less than 14.

The rest of the paper is organized as follows. In Section \ref{chap-buildingd8} we describe how to select a subset of the edges of the Delaunay triangulation $DT(P)$ to form the graph $D8(P)$. In Section \ref{chap-degree} we prove that $D8(P)$ has a maximum degree of 8. In Section \ref{chap-spanner} we bound the spanning ratio of $D8(P)$ with respect to $DT(P)$. Since $DT(P)$ is a spanner of the complete Euclidean graph, this makes $D8(P)$ a spanner of the complete Euclidean graph as well. 

\section{Building D8(P)}\label{chap-buildingd8}

Given as input a set $P$ of $n$ points in the plane, we present an algorithm for building a bounded degree plane graph with maximum degree 8 and spanning ratio bounded by a constant, which we denote as $D8(P)$. The graph denoted $D8(P)$ is constructed by taking a subset of the edges of the Delaunay triangulation of $P$, denoted $DT(P)$.

We assume general position of $P$; i.e., no three points are on a line, no four points are on a circle, and no two points form a line with slope $0$, $\sqrt{3}$ or $-\sqrt{3}$.

\begin{figure}%
	\vspace{-0.5cm}
	\centering
	\subfloat[Cones and bisector distance.] {{	\includegraphics[width = 4cm]{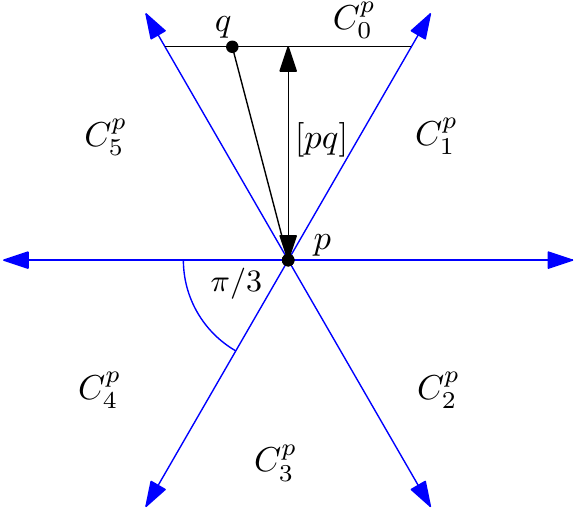} }\label {fig-base-cones}}%
	\qquad
	\subfloat[The \nc $N_i^p$ in black. The thick black lines are canonical edges.] {{\includegraphics [page=1, width = 4cm] {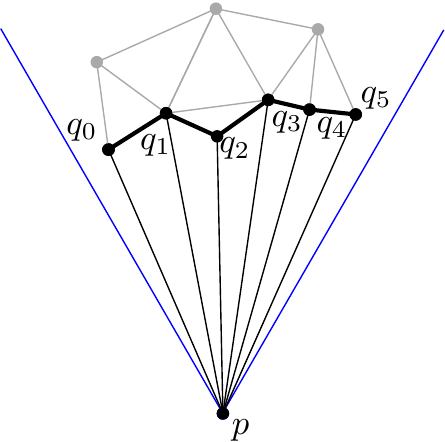} }\label {fig-con-neighbour}}%
	\caption{Preliminaries.}
	\vspace{-0.5cm}
\end{figure}

The space around each vertex $p$ is partitioned by \emph{cones} consisting of 6 equally spaced rays from $p$. Thus each cone has an angle of $\pi/3$. See Figure \ref{fig-base-cones}. We number the cones starting with the topmost cone as $C_0$, then number in the clockwise direction. Cone arithmetic is modulo 6. By our general position assumption we note that no point of $P$ lies on the boundary of a cone.

We introduce a distance function known as the \emph{bisector distance}, which is the distance from $p$ to the orthogonal projection of $q$ onto the bisector of $C_i^p$, where $q\in C_i^p$. We denote this length $[pq]$. Any reference made to distance is to the bisector distance, unless otherwise stated.

\begin{definition}
	Let $\{q_0,q_1,...,q_{d-1}\}$ be the sequence of all neighbours of $p$ in $DT(P)$ in consecutive clockwise order. The neighbourhood $N_p$, with \emph{apex} $p$, is the graph with the vertex set $\{p,q_0,q_1,...,q_{d-1}\}$ and the edge set $\{(p,q_j)\}\cup \{(q_j,q_{j+1})\}, 0 \leq j \leq d-1$, with all values mod $d$. The edges $\{(q_j,q_{j+1})\}$ are called \emph{canonical edges}. $N_i^p$ is the subgraph of $N_p$ induced by all the vertices of $N_p$ in $C_i^p$, including $p$. This is called the \emph{\textbf{\ncp}} of $p$. See Figure \ref{fig-con-neighbour}.
\end{definition}

The algorithm $ConstructD8(P)$ takes as input a point set $P$ and returns the bounded degree graph $D8(P)$, with vertex set $P$ and edge set $E$. The algorithm calls two subroutines. $Add\-Incident()$ selects a set of edges $\EA$. For each edge $(p,r)$ of $\EA$, we call $Add\-Canonical(p,r)$ and $Add\-Canonical(r,p)$ which add edges to the set $\EB$. Both $\EA$ and $\EB$ are a subset of the edges in $DT(P)$. The final graph $D8(P)$ consists of the vertex set $P$ and the union of edge sets $\EA$ and $\EB$.

\begin{wrapfigure}[16]{r}{4cm}
	\begin{center}
		\includegraphics[width = 4cm]{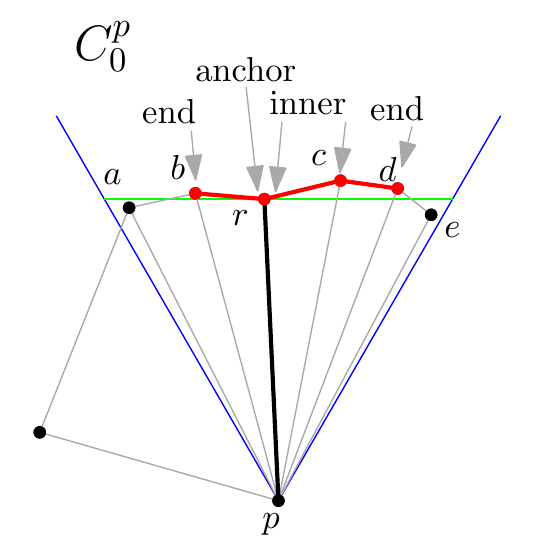}
		\caption{The graph $\C_0^p$, based on $(p,r)\in \EA$, in red. Vertex $r$ is the anchor, $d$ and $b$ are end vertices, and $c$ and $r$ are inner vertices.}\label{fig-canonical-edges}	
	\end{center}
\end{wrapfigure}

We present the algorithm here:\\

 \begin{tabularx}{\textwidth}{l Y}
 \textbf{Algorithm:} & \textbf{ConstructD8(P)}\\
 \textbf{INPUT:} & Set $P$ of $n$ points in the plane.\\
 \textbf{OUTPUT:} & $D8(P)$: spanning subgraph of $DT(P)$.\\
 \end{tabularx}
\begin{enumerate}[labelindent=*,
	style=multiline,
	leftmargin=*,label=Step \arabic*:, ref =Step \arabic*]
\item Compute the Delaunay triangulation $DT(P)$ of the point set $P$. 

\item Sort all the edges of $DT(P)$ by their bisector length, into a set $L$, in non-decreasing order. 

\item Call the function $AddIncident(L)$ with $L$ as the argument. $AddIncident()$ selects and returns the subset $\EA$ of the edges of $L$. 

\item For each edge $(p,r)$ in $\EA$ in sorted order call $AddCanonical(p,r)$ and $Add\-Canonical(r,p)$, which add edges to the set $\EB$.

\item Return $D8(P) = (P, \EA \cup \EB)$. 
\end{enumerate}

 \begin{tabularx}{\textwidth}{l Y}	
\textbf{Algorithm:} & \textbf{AddIncident(L)}\\
\textbf{INPUT:} & $L$: set of edges of $DT(P)$ sorted by bisector distance.\\
\textbf{OUTPUT:} & $\EA$: a subset of edges of $DT(P)$.\\
\end{tabularx}

\begin{enumerate}[labelindent=*,
	style=multiline,
	leftmargin=*,label=Step \arabic*:, ref =Step \arabic*]
	\item Initialize the set $\EA = \emptyset$.
	\item For each $(p,q) \in L$, in non-decreasing order, do:
	\begin{enumerate}
		\item \label{step2a} Let $i$ be the cone of $p$ containing $q$. If $\EA$ has no edges with endpoint $p$ in $N_i^p$, and if $\EA$ has no edges with endpoint $q$ in $N_{i+3}^q$, then we add $(p,q)$ to $\EA$.
	\end{enumerate}
	\item return $\EA$.
\end{enumerate}

The next algorithm requires the following definition:

\begin{definition}\label{pointfive}
	Let $\C_i^{(p,r)}$ be the subgraph of $DT(P)$ consisting of the ordered subsequence of canonical edges $(s,t)$ of $N_i^p$ in clockwise order around apex $p$ such that $[ps]\geq[pr] \text{ and }[pt]\geq[pr]$. We call $\C_i^{(p,r)}$ a \emph{\canp}. A vertex that is the first or last vertex of $\C_i^{(p,r)}$ is called an \emph{end vertex} of $\C_i^{(p,r)}$. A vertex that is not the first or last vertex in $\C_i^{(p,r)}$  is called an \emph{inner vertex} of $\C_i^{(p,r)}$. Vertex $r$ is called the \emph{anchor} of $\C_i^{(p,r)}$. See Fig. \ref{fig-canonical-edges}.
\end{definition}

\begin{figure}%
	\centering
	\subfloat[Edges added to $\EB$ in \ref{unos}.]{{\includegraphics[width=4cm, page=3]{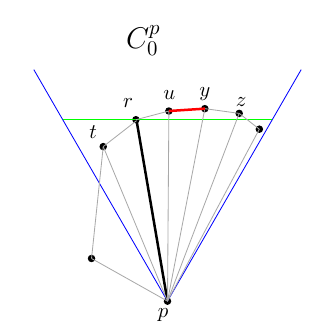} }\label{fig-addcanonical}}%
	\qquad
	\subfloat[Edge added to $\EB$ in \ref{trois}.]{{\includegraphics[width=4cm, page=8]{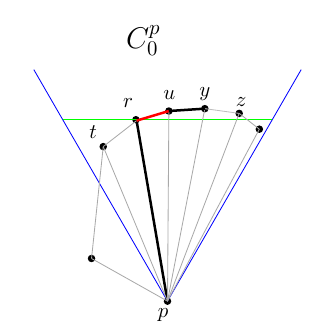} }\label{step3}}%
	\qquad
	\subfloat[Edge added to $\EB$ in \ref{n5}.] {{\includegraphics [width=4cm,page=4] {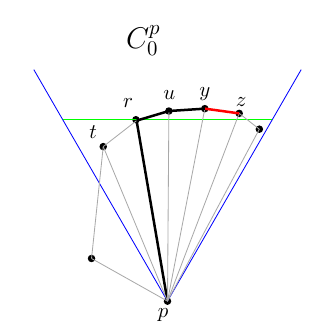} }\label{fig-handlelast}}%
	\qquad
	\subfloat[Edge added to $\EB$ in \ref{aux-edge}.] {{\includegraphics [width=4cm,page=5] {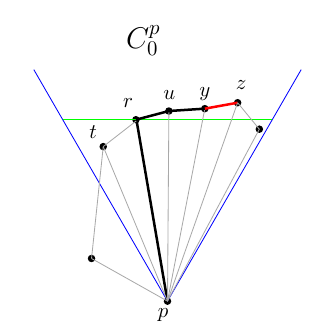} }\label{fig-handlelast2}}%
	\qquad
	\subfloat[Edge added to $\EB$ in \ref{aux-path}.] {{\includegraphics [width=4cm,page=7] {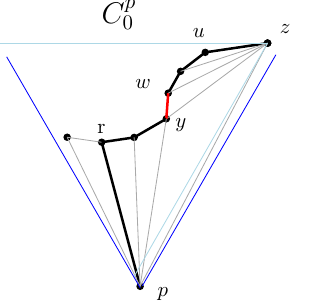} }\label{fig-handlelast4}}%
	\caption{$AddCanonical(p,r)$}%
\end{figure}

\begin{tabularx}{\textwidth}{l Y}
\textbf{Algorithm:}& \textbf{AddCanonical(p,r)}\\
\textbf{INPUT:}& $(p,r)$, an edge of $\EA$.\\
\textbf{OUTPUT:}& A set of edges that are a subset of the edges of $DT(P)$. All edges generated by calls to $AddCanonical()$ form the set $\EB$.\\
\end{tabularx}
\begin{enumerate}[labelindent=*,
	style=multiline,
	leftmargin=*,label=Step \arabic*:, ref =Step \arabic*]
	
	\item Without loss of generality, let $r\in C_0^p$. 
	
	\item \label{unos} If there are at least three edges in $\C_0^{(p,r)}$, then for every canonical edge $(s,t)$ in $\C_0^{(p,r)}$ that is not the first or last edge in the ordered subsequence of canonical edges $\C_0^{(p,r)}$, we add $(s,t)$ to $\EB$.
	
	\item \label{trois} If the anchor $r$ is the first or last vertex in $\C_0^{(p,r)}$, and there is more than one edge in $\C_0^{(p,r)}$, then add the edge of $\C_0^{(p,r)}$ with endpoint $r$ to $\EB$. See Fig. \ref{step3}.
    
    \item\label{step-duos} Consider the first and last canonical edge in $\C_0^{(p,r)}$. Since the conditions for the first and last canonical edge are symmetric, we only describe how to process the last canonical edge $(y,z)$. There are three possibilities. 
	
	\begin{enumerate}
		\item \label{n5} If $(y,z)\in N_5^z$ we add $(y,z)$ to $\EB$. See Fig. \ref{fig-handlelast}.
		
		\item \label{aux-edge}If $(y,z)\in N_4^z$ and $N_4^z$ does not have an edge with endpoint $z$ in $\EA$, then we add $(y,z)$ to $\EB$. See Fig. \ref{fig-handlelast2}
		
		\item \label{aux-path} If $(y,z)\in N_4^z$ and there is an edge with endpoint $z$ in $\EA \cap N_4^z\backslash(y,z)$, then there is exactly one canonical edge of $z$ with endpoint $y$ in $N_4^z$. We label this edge $(w,y)$ and add it to $\EB$. See Fig. \ref{fig-handlelast4}.
	\end{enumerate}
\end{enumerate}

\begin{figure}%
	\centering
	\subfloat[Since $(t,u)$ shown in bold is in $\EA$, and both $u$ and $p$ are in $C_3^t$, $(p,t)$ is not added to $\EA$.]{{\includegraphics[width=4cm]{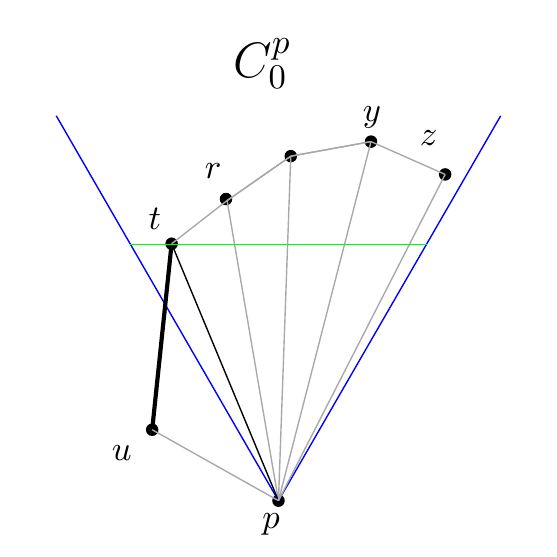} }\label{fig-addincident}}%
	\qquad
	\subfloat[$Add\-Incident()$ proceeds to examine $(p,r)$, which is added to $\EA$.] {{\includegraphics [width=4cm,page=2] {pics/bisector-construction.pdf} }\label{fig-addincident2}}%
	\caption{$AddIncident(L)$ selects edge $(p,r)$ for $\EA$.}
\end{figure}

\section{D8(P) has Maximum Degree 8}\label{chap-degree}

To prove $D8(P)$ has a maximum degree of 8 we use a simple charging scheme. We charge each edge $(p,q)$ of $D8(P)$ once to $p$ and once to $q$. Thus the total charge on a vertex is equal to the degree of that vertex. To help track the number of charges on a vertex, each charge is associated with a specific cone, which may not be the cone containing the edge. We show that a cone can be charged at most twice, and that for any vertex $p$ of $P$, at most two cones of $p$ can be charged twice, while the remaining cones are charged at most once, which yields our maximum degree of 8. 

Sections \ref{cone-section}, \ref{neighbourhood-section}, \ref{shared-section}, and \ref{empty-section} identify the different types of cones and their properties. Sections \ref{charge-ea} and \ref{charge-eb} detail the charging scheme. Section \ref{deg-prove} proves the maximum degree of $D8(P)$. 

\subsection{Cone Types}\label{cone-section}

\begin{figure}%
	\centering
	\subfloat[The region of $N_0^p$ is light blue.]{{\includegraphics[width=4cm, page=5]{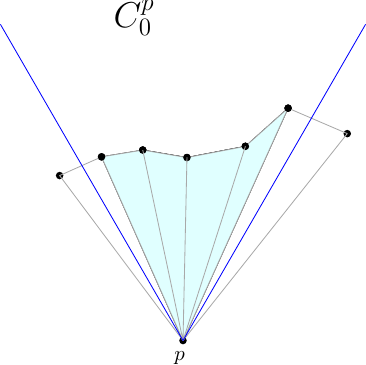} }\label{fig-region2}}%
	\qquad
	\subfloat[Illustrating different types of cones on a canonical region.] {{\includegraphics [width=4cm,page=3] {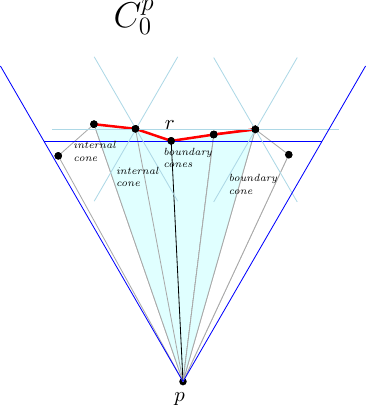} }\label{fig-conetypes}}%
	\caption{Regions and cones.}
\end{figure}

\begin{definition}
	For an arbitrary \nc $N_i^p$ we define the \emph{\rcan of $N_i^p$} as the polygonal region bounded by the canonical edges of $N_i^p$ and the first and last edge of $N_i^p$ with endpoint $p$. See Figure \ref{fig-region2}.
\end{definition}

\begin{definition}
	Let $(a,b)$ be the first edge and let $(y,z)$ be the last edge in a \can $\C_i^{(p,r)}$ (Definition \ref{pointfive}). We define the \emph{\rcan of $\C_i^{(p,r)}$} as the polygonal region bounded by the canonical edges of $N_i^p$ between $(a,b)$ and $(y,z)$ inclusive, and the edges $(p,a)$ and $(p,z)$. See Figure \ref{fig-conetypes}.
\end{definition}

We provide the following definitions regarding the placement of cones in regions. Both of the following definitions also extend to regions of a \ncp.

\begin{definition}
	Let $B(s,\epsilon)$ be a ball with center $s$ and radius $\epsilon>0$. Consider a cone $C_j^s$ of a point $s$ in $\C_i^{(p,r)}$. If there exists an $\epsilon >0$ such that $B(s,\epsilon) \cap C_j^s$ is inside the region of $\C_i^{(p,r)}$, then we call this an \emph{internal cone} of $\C_i^{(p,r)}$. Alternatively we say $C_j^s$ is in $\C_i^{(p,r)}$.
\end{definition}

\begin{definition}
	Consider a cone $C_j^s$ of a point $s$ in $\C_i^{(p,r)}$. If for all $\epsilon >0$, $B(s,\epsilon) \cap C_j^s$ is partially but not entirely in the region of $\C_i^{(p,r)}$, then we call this a \emph{boundary cone} of $\C_i^{(p,r)}$. Alternatively we say $C_j^s$ is on the boundary of $\C_i^{(p,r)}$.
\end{definition}

\begin{definition}
	A cone with vertex $s$ as endpoint is \emph{empty} if no edge of $\EA$ or $\EB$ incident to $s$ is in the cone. 
\end{definition}

\subsection{Cones in Neighbourhoods}\label{neighbourhood-section}

When referring to an angle formed by three points, we refer to the smaller of the two angles (that is, the angle that is $< \pi$) unless otherwise stated. When referring to a circle through three points $p_1,p_2,$ and $p_3$, we use the notation $O_{p_1,p_2,p_3}$.

We consider the edge $(p,r)$ of $\EA$, where without loss of generality, $r$ is in $C_0^p$. In this section we show the location of cones in the \rcan of $\C_0^{(p,r)}$, so we may charge edges of $\EB$ to them. To facilitate this we introduce another variation on the concept of the neighbourhood of a vertex:

\begin{definition}
	Consider the \nc $N_i^p$ with the vertex set $\{p,q_0,q_1,$ $... ,q_{m-1}\}$, where $\{q_0,q_1, ... ,q_{m-1}\}$ are listed in clockwise order around $p$. A \emph{\rn} $N_p^{(q_j,q_k)}$ is the subgraph of $N_i^p$ induced on the vertex set $\{p,q_j,q_{j+1} ... ,q_k\}, 0 \leq j \leq k \leq m-1$. 
\end{definition}

Now we illustrate some of the geometric properties of \rns in $DT(P)$.

\begin{lemma}\label{lemma-inside-edge-in-disk}
	Consider the arbitrary \rn $N_p^{(r,q)}$. Each vertex $x \in  N_p^{(r,q)}\backslash\{p,r,q\}$ is in the circle $O_{p,r,q}$ through $p$, $r$, and $q$.
\end{lemma}

\begin{figure}%
	\centering
	\subfloat[When corners of a quadrilateral lie on a circle, opposite angles will sum to $\pi$.]{{\includegraphics[width=4cm, page=1]{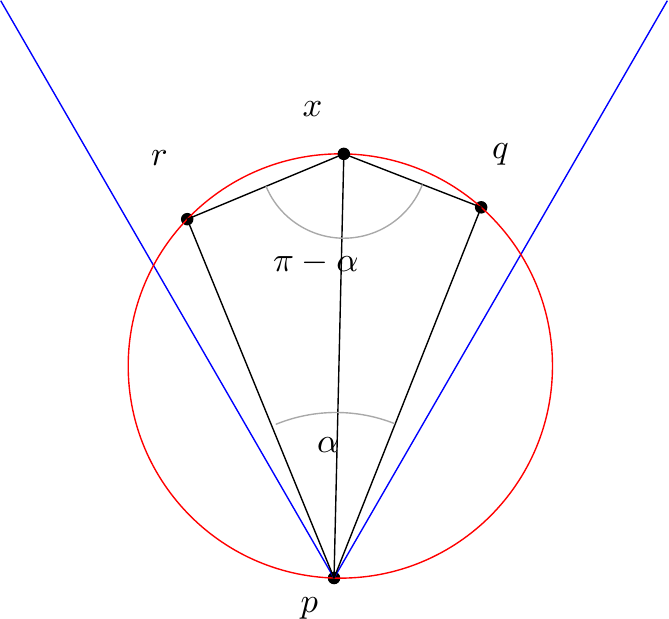} }\label{circ1}}%
	\qquad
	\subfloat[When one corner of a quadrilateral lies inside a circle, that corner and its opposite angles will sum to $>\pi$.] {{\includegraphics [width=4cm,page=2] {pics/circle-inside-edge.pdf} }\label{circ2}}%
	\qquad
	\subfloat[When one corner of a quadrilateral lies outside a circle, that corner and its opposite angle will sum to $<\pi$.] {{\includegraphics [width=4cm,page=3] {pics/circle-inside-edge.pdf} }\label{circ3}}%
	\caption{Properties of convex quadrilaterals in $DT(P)$.}
\end{figure}

\begin{proof}
	Since $(p,x)$ is an edge in $DT(P)$,  we can draw a disk through $p$ and $x$ that is empty of points of $P$. In particular, neither $r$ nor $q$ is in this disk. Hence the sum of the angles $\angle(prx)$ and $\angle(pqx)$ which lie on opposite sides of the same chord is smaller than $\pi$, and the sum of the other two angles $\angle(rxq)$ and $\angle(rpq)$ in the quadrilateral $(prxq)$ is greater then $\pi$. That implies $x$ is inside $O_{p,r,q}$. 
\end{proof}

\begin{lemma}\label{lemma-wedge-opposite-angles}
	
	Consider the \rn $N_p^{(r,q)}$ in cone $C_i^p$. Let $(p,x)$ be an edge in $N_p^{(r,q)}$ where $x \neq r$ and $x \neq q$. Then angle $\angle(qxr) \geq \pi - \angle(qpr)$. Since the cone angle is $\pi/3$, we have that $\angle(qxr)>2\pi/3$.
\end{lemma}

\begin{proof}
	We know by Lemma \ref{lemma-inside-edge-in-disk} that $x$ lies inside the circle through $p$, $r$ and $q$, which we label $O_{p,r,q}$. The angle $\angle(qxr)$ is minimized when $x$ is on $O_{p,r,q}$. When $x$ is on $O_{p,r,q}$, $\angle rxq = \pi - \angle(qpr)$, since the two angles lie on the same chord $(r,q)$. Therefore $\angle(rxq) \geq \pi - \angle(qpr)$. Since both $q$ and $r$ are in the same cone $C_i^p$, and the cone angle is $\pi/3$, the $\angle(qxr)>2\pi/3$. 
\end{proof}

Which leads to the corollary:
\begin{corollary} \label{cor-empty-cone}
	Let $s$ be an inner vertex of $\C_i^{(p,r)}$ that is not the anchor. Then there is at least one empty cone of $s$ in $\C_i^{(p,r)}$. 
\end{corollary}

\begin{figure}
	\centering
	\subfloat[Properties of convex quadrilaterals in $DT(P)$.]{\includegraphics[width = 4cm, page = 4]{pics/circle-inside-edge.pdf}\label{circ}}
	\qquad
	\subfloat[Consecutive canonical edges have an angle facing $p$ of at least $2\pi/3$, and thus if $(p,r)\notin \EA$, there is at least one empty cone between them in $D8(P)$.]{
		\includegraphics[page=1, width = 4cm]{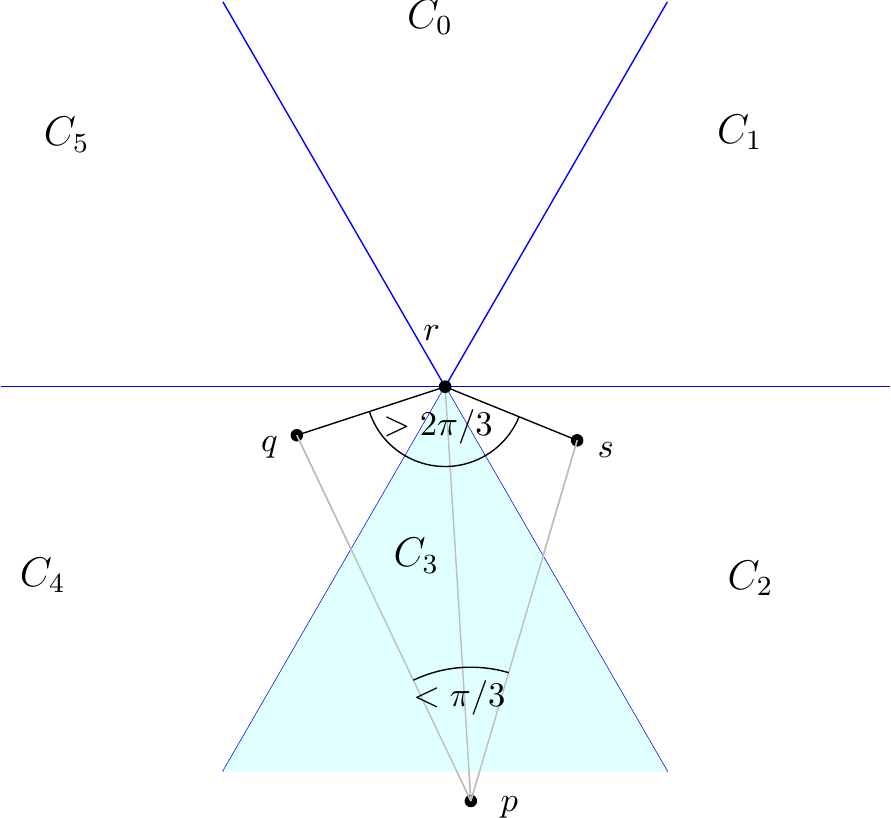}\label{fig-can-charging}}
	\qquad
	\subfloat[Lemma \ref{r-edges}.]{\includegraphics[page=1, width = 4cm]{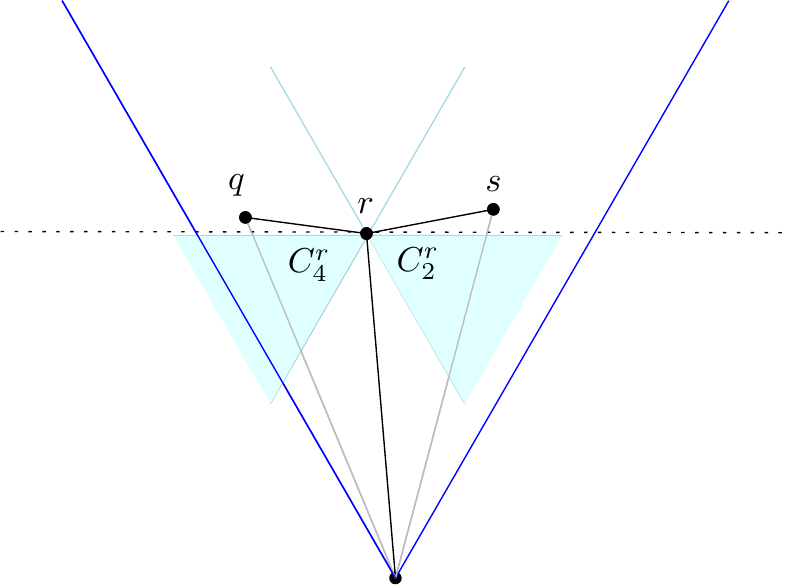}\label{fig-r-edges}}\caption{Locating empty cones.}
	
\end{figure}

\begin{proof}
	Since $s$ is not the anchor, any internal cone of $\C_i^{(p,r)}$ on vertex $s$ is empty, and by Lemma \ref{lemma-wedge-opposite-angles}, there is at least one internal cone of $\C_i^{(p,r)}$ on vertex $s$. Therefore there is at least one empty internal cone on $s$ in the region of $\C_i^{(p,r)}$. See Fig. \ref{fig-can-charging}.
	\qed \end{proof}

\begin{lemma}\label{r-edges}
	Consider the edge $(p,r)$ in $\EA$, and without loss of generality let $r$ be in $C_0^p$. If $r$ is an inner anchor of $\C_0^{(p,r)}$, then cones $C_2^r$ and $C_4^r$ are empty and in the region of $\C_0^{(p,r)}$. If $r$ is an end vertex and not the only vertex in $\C_0^{(p,r)}$, then at least one of $C_2^r$ and $C_4^r$ are empty and in the region of $\C_0^{(p,r)}$.
\end{lemma}

\begin{proof}
	Since $(p,r)$ is in $C_0^p$, it must also be in $C_3^r$, and thus it is in neither $C_2^r$ or $C_4^r$.
	
	If $r$ is an inner vertex, assume that $q$, $r$ and $s$ are in consecutive order in $\C_0^{(p,r)}$. Thus $\C_i^{(p,r)}$ contains canonical edges $(q,r)$ and $(r,s)$. 
	
	Recall that for every vertex $x$ in $\C_0^{(p,r)}$, $[px]\geq[pr]$. Thus $[ps]\geq[pr]$ and $[pq]\geq[pr]$, which means that $s$ and $q$ are above the horizontal line through $r$ in $C_0^p$. Since $C_2^r$ and $C_4^r$ lie below the horizontal line through $r$, they cannot contain the edges $(q,r)$ and $(r,s)$. Since $\triangle(pqr)$ and $\triangle(prs)$ are triangles in $DT(P)$, $C_2^r$ and $C_4^r$ are empty and inside $\C_0^{(p,r)}$. See Fig. \ref{fig-r-edges}.
	
	Otherwise $r$ is an end vertex. By the same argument as above, but applied to only one side of $(p,r)$, either $C_2^r$ or $C_4^r$ is empty. 
	\qed \end{proof}

\subsection{Cones in Shared Triangles}\label{shared-section} 

We will show the location of uncharged cones in the special case of overlapping \rcansp.
A set of regions overlap when at least one triangle of $DT(P)$ is contained in the intersection of all the regions in the set.  

\begin{lemma}\label{awesome_lemma}
	Consider the triangle $\triangle(pp's)$ in $DT(P)$. Let $p'$ and $s$ be in $C_0^p$, and let $p$ and $s$ be in $C_3^{p'}$. Then $\angle(p'sp)>2\pi/3$. 
\end{lemma}

\begin{figure}
	\centering
	\includegraphics[page=3, width = 6cm]{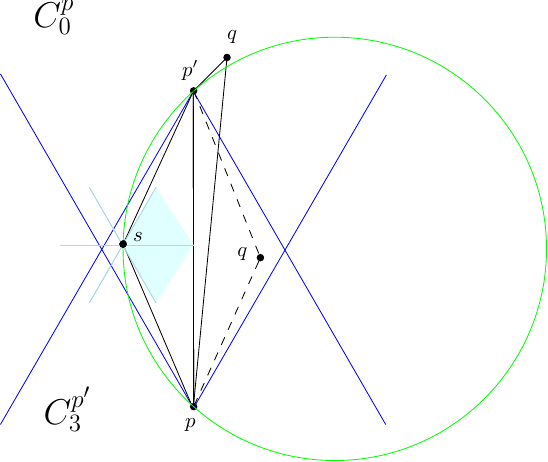}\caption{$q \in C_0^p \cap C_3^{p'}$ violates the empty circle property of Delaunay triangulations.}\label{fig-sharedtriangle}
\end{figure}

\begin{proof}
	Without loss of generality, assume that $s$ is left of directed line segment $(p,p')$.	Consider the parallelogram formed by $C_0^p\cap C_3^{p'}$. Let $a$ be the left intersection and $b$ be the right intersection of $C_0^p$ and $C_3^{p'}$. Thus $s$ is in $\triangle(app')$. Note that $\angle(pp'a) + \angle(app') = \pi/3$. Thus $\angle(pp's) + \angle(spp') < \pi/3$, which implies that $\angle(p'sp)>2\pi/3$. See Figure \ref{fig-sharedtriangle}.
\end{proof}

\begin{lemma}\label{at-most-two}
	Let $\triangle(pp's)$ be a triangle in $DT(P)$. Then $\triangle(pp's)$ can belong to \ncs of at most two of $p$, $p'$ and $s$. 
\end{lemma}

\begin{proof}
	Suppose, for the sake of contradiction, that triangle $\triangle(pp's)$ is in a \nc of $p$, and a \nc of $p'$, and a \nc of $s$. Without loss of generality, let $p'$ and $s$ be in $C_0^p$. This means that $p'$ and $s$ must be in $C_3^{p'}$. By Lemma \ref{awesome_lemma} angle $\angle(psp')>2\pi/3$. Therefore $p$ and $p'$ cannot be in the same \nc of $s$.
\end{proof}

\begin{corollary}\label{cor-shared}
	A triangle can be shared by at most 2 \ncsp.
\end{corollary}

\begin{proof}
	Follows from Lemma \ref{at-most-two}.
\end{proof}

Which leads to the following definition:

\begin{definition}
	If $\triangle(pp's)$ occurs in exactly \emph{two} \ncs of $p$,$p'$ and $s$, then we refer to it as a \emph{shared triangle}. If $\triangle(pp's)$ is in \ncs of $p$ and $p'$, then $(p,p')$ is referred to as the \emph{base} of the shared triangle. 
\end{definition}

\begin{corollary}\label{shared-empty-cones}
	In a shared triangle $\triangle(pp's)$ with base $(p,p')$, $s$ has two empty cones internal to $\triangle(pp's)$.
\end{corollary}

\begin{proof}
	Without loss of generality, assume that $p' \in C_0^p$ and $p \in C_3^{p'}$. Then $p \in C_3^s$ and $p' \in C_0^s$. Then either $C_2^s$ and $C_1^s$, or $C_5^s$ and $C_4^s$ are internal to $\triangle(pp's)$ and thus cannot contain any edges with endpoint $s$. 
\end{proof}

We show two \ncs can share at most one triangle:

\begin{lemma}\label{lemma-two-shared-triangles}
	Each Delaunay edge is the base of at most 1 shared triangle.
\end{lemma}

\begin{proof}
	By contradiction.
	
	Consider the shared triangle $\triangle(pp's)$ with base $(p,p')$. Without loss of generality, assume $p'$ is in $C_0^p$, and $p$ is in $C_3^{p'}$. Since $(p,p')$ is an edge in exactly two triangles, let $\triangle(pp'q)$ be the other triangle with edge $(p,p')$, and assume that $q$ is in both $C_0^p$ and $C_3^{p'}$.
	
	By Lemma \ref{awesome_lemma} both angles $\angle(p'sp)$ and $\angle(pqp')$ are greater than $2\pi/3$. Thus their sum is greater than $4\pi/3$. But by the empty circle property of Delaunay triangulations,  $\angle(p'sp) + \angle(pqp')$ must be less than $\pi$, which is a contradiction. 
\end{proof}

\begin{corollary}
	Two \ncs can share at most one triangle in $DT(P)$.
\end{corollary}

\subsection{Empty Cones}\label{empty-section}

\begin{lemma}\label{lemma-empty-cone}
	Consider the edge $(p,r)$ in $\EA$. Without loss of generality let $r$ be in $C_0^p$. Each inner vertex $s$ of $\C_0^{(p,r)}$ that is not the anchor has at least one unique empty cone in the region of $\C_0^{(p,r)}$.
\end{lemma}

\begin{proof}
	
	If $s$ is not part of a shared triangle, we know by Corollary 	\ref{cor-empty-cone} that $s$ has an empty cone internal to $\C_0^{(p,r)}$.
	
	Consider the shared triangle $\triangle(pp's)$, and without loss of generality, let $p'$ be in $C_0^p$. Assume that there is an edge $(p,r)$ of $\EA$ in $C_0^p$, and an edge $(p',r')$ of $\EA$ in $C_3^{p'}$. Thus both sets $\C_0^{(p,r)}$ and $\C_3^{(p',r')}$ are well-defined. 
	
	By Corollary \ref{shared-empty-cones} there are two empty cones of $s$ internal to $\triangle(pp's)$. The empty cone adjacent to $(p',s)$ is the empty cone of $s$ in the region of $\C_0^{(p,r)}$, and the empty cone adjacent to $(p,s)$ is the empty cone of $s$ in the region of $\C_3^{(p',r')}$. 
	
	Thus any inner vertex $s$ of an arbitrary \can $\C_0^{(p,r)}$ that is not the anchor, has a unique empty cone that is in the region of $\C_0^{(p,r)}$.\\
\end{proof}

\begin{lemma}\label{lemma-empty-cone-end}
	Consider the edge $(p,r)$ in $\EA$. Without loss of generality let $r$ be in $C_0^p$. Let $z \neq r$ be an end vertex in $\C_0^{(p,r)}$. By symmetry, let $z$ be the last vertex. Let $y$ be the neighbour of $z$ in $\C_0^{(p,r)}$.  If $y$ is in $C_5^z$, then $C_4^z$ is a unique empty cone internal to the region of $\C_0^{(p,r)}$.
\end{lemma}

\begin{proof}
	Triangle $\triangle(pyz)$ is a triangle in $DT(P)$. Since $(p,z)$ is in $C_3^z$ and $(y,z)$ is in $C_5^z$, $C_4^z$ will have no edges in $DT(P)$ with endpoint $z$. Since both $\EA$ and $\EB$ are subsets of the edges of $DT(P)$, $C_4^z$ will not contain any edges of $\EA$ and $\EB$ with endpoint $s$, and thus is empty. 
	
	We prove $C_4^z$ is unique to $\C_0^{(p,r)}$ by contradiction. Since $C_4^z$ is inside a triangle of $DT(P)$, it cannot be a boundary cone, thus it must be inside of shared triangle $\triangle(pyz)$. Corollary \ref{shared-empty-cones} states that $\triangle(pyz)$ must have two empty cones internal to $\triangle(pyz)$. However, since $(p,z)$ is in $C_3^z$, and $(y,z)$ is in $C_5^z$, only $C_4^z$ is an empty internal cone of $\triangle(pyz)$, which is a contradiction.
\end{proof}

\begin{lemma}\label{lemma-empty-cone-anchor}
	Consider the edge $(p,r)$ in $\EA$, and without loss of generality let $r$ be in $C_0^p$ (thus $r$ is the anchor of $\C_0^{(p,r)}$). The empty cones of $r$ internal to $\C_0^{(p,r)}$ are unique to $\C_0^{(p,r)}$.
\end{lemma}

\begin{proof}
	By Lemma \ref{r-edges}, $C_2^r$ and $C_4^r$ are (possibly) empty cones inside $\C_0^{(p,r)}$. Since the cases are symmetric, we consider $C_2^r$. Assume $s$ is the neighbour of $r$ in $\C_0^{(p,r)}$ such that $C_2^r$ is inside $\triangle(rsp)$. 
	
	If $(r,s)$ is in $C_1^r$, then $C_2^r$ is the only empty cone inside $\triangle(rsp)$. By Corollary \ref{shared-empty-cones} a shared triangle must have two empty cones, thus $C_2^r$ must be unique to $\C_0^{(p,r)}$.
	
	Otherwise, if $(r,s)$ is in $C_0^r$, then $\triangle(rsp)$ is a shared triangle. Since both  $C_2^r$ and $C_1^r$ are empty, we designate $C_2^r$ as belonging to $\C_0^{(p,r)}$, and $C_1^r$ as belonging to $\C_3^{(s,\cdot)}$. Thus $C_2^r$ is unique to $\C_0^{(p,r)}$
\end{proof}

\subsection{Charging Edges in $\EA$}\label{charge-ea}

The charging scheme for the edges of $\EA$ is as follows. Consider an edge $(p,r)$ of $\EA$, where without loss of generality $r$ is in $C_0^p$ and $p$ is in $C_3^r$. An edge $(p,r)$ of $\EA$ charges $C_0^p$ once and $C_3^r$ once.

\begin{lemma}\label{lemma-deg6}
	Each cone of an arbitrary vertex $p$ of the graph $D8(P)$ is charged at most once by an edge of $\EA$ (thus yielding a maximum degree for the graph $G=(P,\EA)$ of 6). 
\end{lemma}

\subsection{Charging Edges in $\EB$}\label{charge-eb}

Let $(p,r)$ be an edge of $\EA$, and without loss of generality let  $r\in C_0^p$.  Let $\C_0^{(p,r)}$ be the subgraph consisting of the ordered subsequence of canonical edges $(s,t)$ of $N_0^p$ in clockwise order around apex $p$ such that $[ps]\geq[pr] \text{ and }[pt]\geq[pr]$. We call $\C_0^{(p,r)}$ a \canp.

For edges in $\EB$ we consider an arbitrary \can $\C_i^{(p,r)}$, and without loss of generality let $i=0$. We note that there are three types of vertices in $\C_0^{(p,r)}$: anchor, inner and end vertices. Thus any edge added to $\EB$ from $\C_0^{(p,r)}$ will be charged to an inner, end or anchor vertex (refer to Fig \ref{fig-canonical-edges}). We outline the charging scheme below by referencing the steps of $AddCanonical(p,r)$ where edges were added to $\EB$.

\begin{figure}%
	\centering
	\subfloat[Vertex $r(s)$ is the anchor. Charge $(r,t_1)$ to $C_4^s$ and $(r,t_2)$ to $C_2^s$.] {{\includegraphics[page=18, width = 3.5cm]{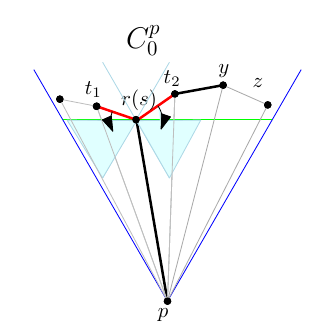} }\label{ttq1}}%
	\qquad
	\subfloat[Vertex $r(s)$ is the anchor. Charge $(r,t)$ to $C_2^s$.] {{\includegraphics[page=4, width = 3.5cm]{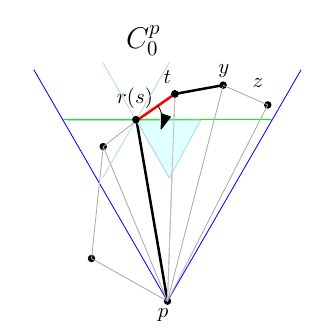} }\label{ttq5}}%
	\qquad
	\subfloat[Vertex $z$ is the last vertex of $\C_0^{(p,r)}$, $z \neq r$.] {{\includegraphics[page=5, width = 3.5cm]{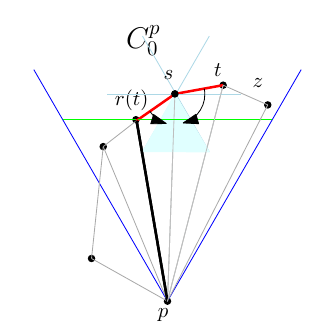} }\label{ttq2}}%
	\qquad
	\subfloat[Vertex $y$ is neighbour to the last vertex $z$.] {{\includegraphics[page=8, width = 3.5cm]{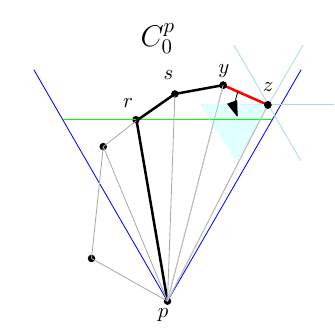}}\label{ttq3}}%
	\qquad
	\subfloat[Vertex $y$ charges both edges to its empty cone.                                                                                                                                                                                                                                                                                                                                                                                                                                                                                                                                                                                                                                                                                                                                                                                                                                                                                                                                                                                                                                                                                                                                                                                                                                                                                                                                                                                                                                                                                                                                                                                                                                                                                                                                                                                                                                                                                                                                                                                                                                                                                                                                                                                                                                                                                                                                                                                                                                                                                                                                                                                                                                                                                                                                                                                                                                                                                                                                                                                                                                                                                                                                                                                                                                                                                                                                                                                                                                                                                                                                                                                                                                                                                                                                                                                                                                                                                                                                                                                                                                                                                                                                                                                                                                                                                                                                                                                                                                                                                                                                                                                                                                                                                                                                                                                                                                                                                                                                                                                                                                                                                                                                                                                                                                                                                                                                                                                                                                                                                                                                                                                                                                                                                                                                                                                                                                                                                                                                                                                                                                                                                                                                                                                                                                                                                                                                                                                                                                                                                                                                                                                                                                                                                                                                                                                                                                                                                                                                                                                                                                                                                                                                                                                                                                                                                                                                                                                                                                                                                                                                                                                                                                                                                                                                                                                                                                                                                                                                                                                                                                                                                                                                                                                                                                                                                                                                                                                                                                                                                                                                                                                                                                                                                                                                                                                                                                                                                                                                                                                                                                                                                                                                                                                                                                                                                                                                                                                                                                                                                                                                                                                                                                                                                                                                                                                                                                                                                                  ] {{\includegraphics[page=9, width = 3.5cm]{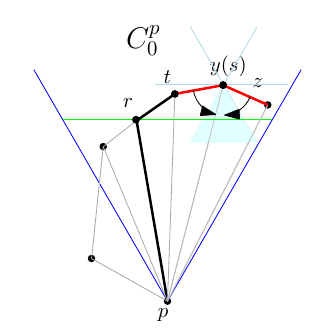}}\label{ttq4}}%
	\qquad
	\subfloat[$(y,z)$ is the last canonical edge in $\C_0^{(p,r)}$, and is charged to $C_4^z$.] {{\includegraphics[page=11, width = 3.5cm]{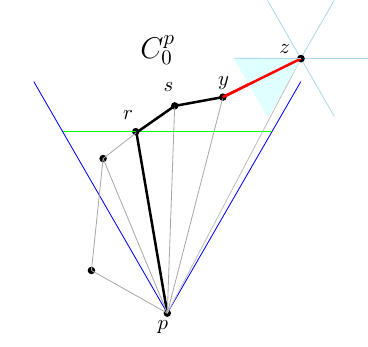}}\label{aux-edge2}}%
	\qquad
	\subfloat[$(y,z)$ is the last canonical edge in $\C_0^{(p,r)}$, and is charged to the empty cone of $y$.] {{\includegraphics[page=12, width = 3.5cm]{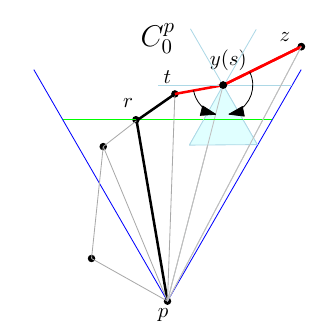}}\label{aux-edge3}}%
	\qquad
	\subfloat[$(w,y)$ is charged to $y$ in place of $(y,z)$.] {{\includegraphics[page=16, width = 3.5cm]{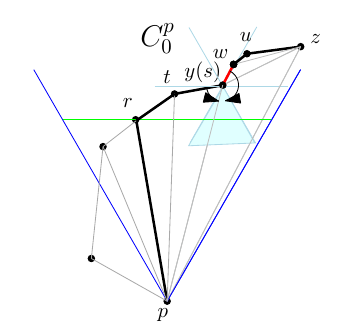}}\label{can-edge2}}%
	\qquad
	\subfloat[$(w,y)$ is charged to the empty cone of $w$.] {{\includegraphics[page=15, width = 3.5cm]{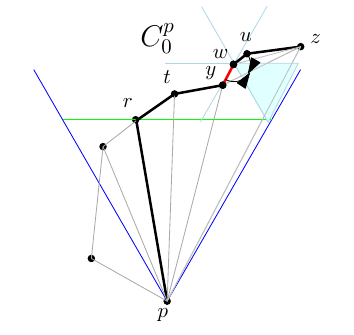}}\label{can-edge3}}%
	\caption{Charging scheme for edges of $\EB$.}
\end{figure}

\begin{enumerate}[labelindent=*,
	style=multiline,
	leftmargin=*,label=Step \arabic*:, ref = Step \arabic*]
	\item Without loss of generality, let $r\in C_0^p$. 
	\item If the anchor $r$ is the first or last vertex in $\C_0^{(p,r)}$, and there is more than one edge in $\C_0^{(p,r)}$, then add the edge of $\C_0^{(p,r)}$ with endpoint $r$ to $\EB$. See Fig. \ref{step3}.
	
	\item \label{unos-c} \emph{If there are at least three edges in $\C_0^{(p,r)}$, then for every canonical edge $(s,t)$ in $\C_0^{(p,r)}$ that is not the first or last edge in the ordered subsequence of canonical edges $\C_0^{(p,r)}$, we add $(s,t)$ to $\EB$.}
	
	The edge $(s,t)$ is charged once to $s$ and once to $t$. Since the charging scheme is the same for both $s$ and $t$, without loss of generality we only describe how to charge $s$.  
	
	\textbf{Charge vertex $s$:} (Steps 1 and 2)
	\begin{enumerate}
		
		\item \label{charge1a} If $s$ is the anchor (thus $s=r$), then by Lemma \ref{r-edges}, $C_2^r$ and $C_4^r$ are empty cones inside $\C_0^{(p,r)}$. If $t$ is left of directed line segment $pr$, charge $(r,t)$ to $C_4^r$. If $t$ is right of $pr$, charge $(r,t)$ to $C_2^r$. See Fig. \ref{ttq4}.
		
		\item \label{charge1b} If $s\neq r$ then by Lemma \ref{lemma-empty-cone}, $s$ has an empty cone $C_j^s$ inside $\C_0^{(p,r)}$. Charge $(s,t)$ once to $C_j^s$. See Fig.s \ref{ttq1}, \ref{ttq2}, \ref{ttq3}.
		
	\end{enumerate}
	
	\item\label{step-duos-c} \emph{Consider the first and last canonical edge in $\C_0^{(p,r)}$. Since the conditions for the first and last canonical edge are symmetric, we only describe how to process the last canonical edge $(y,z)$. There are three possibilities.}
	
	\begin{enumerate}[label = (\alph*), ref = (\alph*)]
		\item \label{n5-c} \emph{If $(y,z)\in C_5^z$, add $(y,z)$ to $\EB$. See Fig. \ref{fig-handlelast}.}
		
		\item \label{aux-edge-c}\emph{If $(y,z)\in N_4^z$ and $N_4^z$ does not have an edge with endpoint $z$ in $\EA$, then we add $(y,z)$ to $\EB$. See Fig. \ref{fig-handlelast2}}
		
		\textbf{Charge vertex $y$:}
		\setcounter{MyCounter}{1}
		\begin{enumerate}[label={\roman{MyCounter}}]
			\item \label{charge2a} If $y$ is the anchor, then $C_2^y$ is empty and inside $\C_0^{(p,r)}$ by Lemma \ref{r-edges}. Charge $(y,z)$ to $C_2^y$. Fig. \ref{ttq4}.
			
			\addtocounter{MyCounter}{1}
			\item \label{charge2b}Otherwise $y$ is not the first or last vertex in $\C_0^{(p,r)}$, and by Corollary \ref{cor-empty-cone} has an empty cone $C_j^y$ inside $\C_0^{(p,r)}$. Charge $(y,z)$ to $C_j^y$. Fig.s \ref{ttq1}, \ref{ttq2}, \ref{ttq3}.
		\end{enumerate}
		
		\textbf{Charge vertex $z$:}
		\begin{enumerate}[label={\roman{MyCounter}}]
			
			\addtocounter{MyCounter}{1}
			\item  \label{charge2c}\ref{n5}: $(y,z)$ is in $C_5^z$. By Lemma \ref{lemma-empty-cone-end} $C_4^z$ is empty and inside $\C_0^{(p,r)}$. Charge $(y,z)$ to $C_4^z$. Fig. \ref{aux-edge3}.
			
			\addtocounter{MyCounter}{1}
			\item \label{charge2d}\ref{aux-edge}: $(y,z)$ is in $C_4^z$, and $C_4^z$ does not contain an edge of $\EA$ with endpoint $z$. Note $C_4^z$ is a boundary cone of $\C_0^{(p,r)}$. Charge $(y,z)$ to $C_4^z$. Fig. \ref{aux-edge2}.
		\end{enumerate}
		
		\item \label{aux-path-c} \emph{If $(y,z)\in C_4^z$ and there is an edge $(u,z), u \neq y$ of $\EA$ in $C_4^z$, then there is one canonical edge of $z$ with endpoint $y$ in $C_4^z$. Label the edge $(w,y)$ and add it to $\EB$. See Fig. \ref{fig-handlelast4}.}
		
		\textbf{Charge vertex $y$:}
		\setcounter{MyCounter}{1}
		\begin{enumerate}[label={\roman{MyCounter}}]
			\item \label{charge3a}If $y=r$, then $C_2^y$ is empty and inside $\C_0^{(p,r)}$ by Lemma \ref{r-edges}. Charge $(w,y)$ to $C_2^y$. Fig. \ref{ttq4}.
			
			\addtocounter{MyCounter}{1}
			\item \label{charge3b}Otherwise $y$ is not the first or last vertex in $\C_0^{(p,r)}$, and by Corollary \ref{cor-empty-cone} has an empty cone $C_j^y$ inside $\C_0^{(p,r)}$. Charge $(w,y)$ to $C_j^y$. Fig. \ref{can-edge2}.
		\end{enumerate}
		
		\textbf{Charge vertex $w$:}
		\begin{enumerate}[label={\roman{MyCounter}}]
			
			\addtocounter{MyCounter}{1}
			\item \label{charge3c} If $w=u$ ($(z,u)$ in $\EA$), then $C_2^w$ is empty and inside $\C_4^{(z,u)}$ by Lemma \ref{empty-cone}. Charge $(w,y)$ to $C_2^w$. Fig. \ref{ttq4}.
			
			\addtocounter{MyCounter}{1}
			\item \label{charge3d}If $w \neq u$, then $w$ is not the first or last vertex in $\C_4^{(z,u)}$, and by Corollary \ref{cor-empty-cone} has an empty cone $C_j^w$ inside $\C_4^{(z,u)}$. Charge $(w,y)$ to $C_j^w$. Fig. \ref{can-edge3}.
		\end{enumerate}
	\end{enumerate}
	
\end{enumerate}

\ref{step-duos-c}\ref{aux-path-c}\ref{charge3c} makes use of the following lemma:\\

\begin{lemma}\label{empty-cone}
	Assume that on a call to $AddCanonical(p,r)$, where $(p,r)$ is in $C_0^p$, we add edge $(w,y)$ to $\EB$ in \ref{aux-path}. Let $(y,z)$ be the last edge in $\C_0^{(p,r)}$, and assume that $(w,z)$ is in $\EA$. Then $C_2^w$ is empty and inside $\C_4^{(z,u)}$. 
\end{lemma}

\begin{figure}
	\centering
	\includegraphics{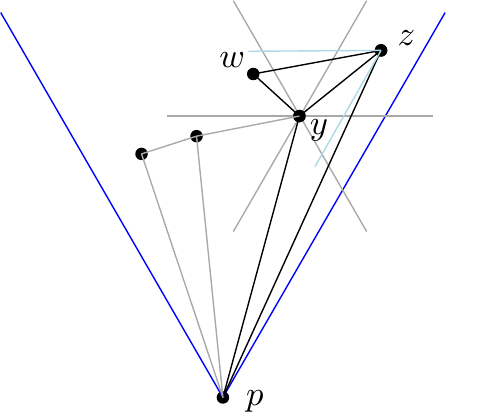}
	\caption{The case if $[wz]>[yz]$.}\label{wyincan}
\end{figure}

\begin{proof}
	To prove this we shall establish that $[yz]\geq[wz]$. This, together with Lemma \ref{r-edges} implies that $C_2^w$ is empty and inside $\C_4^{(z,u)}$. 
	
	We prove by contradiction, thus assume that $[wz]>[yz]$. See Fig. \ref{wyincan}. This means that $AddIncident(L)$ examined $(y,z)$ before $(w,z)$, and thus $C_4^z \cap \EA$ was empty of edges with endpoint $z$ when $(y,z)$ was examined by $AddIncident(L)$. Since $(y,z)$ was not added to $\EA$, $y$ must have had an edge of $\EA$ in $C_1^y$ with endpoint $y$ that was shorter than $(y,z)$.
	
	Since $\triangle(pyz)$ is a triangle in $DT(P)$, and $p \in C_3^y$, there cannot be an edge with endpoint $y$ in $C_1^y$ clockwise from $(y,z)$. In the counter-clockwise direction from $(y,z)$, we have the $\triangle(ywz) \in DT(P)$. However, since $[wz]>[yz]$, $(w,y)$ cannot be in $C_1^y$. Thus $C_1^y$ contained no edge of $\EA$ with endpoint $y$ when $(y,z)$ was examined by $AddIncident(L)$. Which means if $[wz]>[yz]$, then $(y,z)$ would have be added to $\EA$ by $AddIncident(L)$. But we know $(w,z)$ is in $\EA$, therefore it must be that $[yz]\geq[wz]$, which implies that $C_2^w$ is empty and by Lemma \ref{r-edges} is inside $\C_4^{(z,u)}$.
	\qed \end{proof}

\subsection{Proving the Degree of D8(P)}\label{deg-prove}

The charging argument of the previous section establishes where charges are made. In this section we show a limit to how many edges can be charged to the different cones. 


All edges added to $\EB$ are charged to internal cones, with the exception of the edge that is added in $AddCanonical(p,\cdot)$ \ref{aux-edge} to $C_4^z$, which is to a boundary cone. Since $C_4^z$ is on the boundary of $\C_0^{(p,r)}$, it may also be the boundary cone of a different \ncp. 

\begin{lemma}\label{no-internal}
	The boundary cone $C_4^z$ of $\C_0^{(p,r)}$ charged in $AddCanonical(p,r)$ \ref{aux-edge} cannot be the internal cone of a different \can.
\end{lemma}

\begin{proof}
	$C_4^z$ is inside a \can, then $y$ must be the apex of said \canp. That implies that both shared neighbours of $z$ and $y$ must be in $C_1^y$. But shared neighbour $p$ is in $C_3^y$, thus $C_4^z$ cannot be inside a \can.
	\qed \end{proof}

This implies that $C_4^z$ may only be shared with a different \can as a boundary cone. Thus the only other edge of $\EB$ that can be charged to $C_4^z$ must be added in some call to  $AddCanonical(\cdot,\cdot)$ \ref{aux-edge}. We prove here this is impossible.

\begin{lemma}\label{lemma-double-canonical}
	Consider the edge $(p,r)$ of $\EA$ in $C_i^p$, and without loss of generality, let $i=0$. Let $(y,z)$ be the last edge in $\C_0^{(p,r)}$, and let $z$ be the last vertex in $\C_0^{(p,r)}$. Assume that $(y,z)$ was added to $\EB$ in a call to $AddCanonical(p, r)$ \ref{aux-edge}, and thus by Charge \ref{charge2d} is charged to the cone $C_4^z$. Then $(y,z)$ is the only edge in $D8(P)$ charged to $C_4^z$.	
\end{lemma}

\begin{proof}
	
	Edge $(y,z)$ is added to $\EB$ in a call to $AddCanonical(p, r)$ \ref{aux-edge}, only if there is no edge of $\EA$ in $C_4^z$. Thus $C_4^z$ is not charged by an edge of $\EA$.
	
	Cone $C_4^z$ is a boundary cone of $\C_0^{(p,r)}$. By Lemma \ref{no-internal}, $C_4^z$ cannot be an internal cone of another \canp. Thus $C_4^z$ can only be a boundary cone of any \canp. Since $AddCanonical(\cdot, \cdot)$ \ref{aux-edge} is the only call that adds an edge to $\EB$ that is charged to a boundary cone, only another call to $AddCanonical(\cdot, \cdot)$ \ref{aux-edge} can charge an additional edge to $C_4^z$.
	
	Assume we have edges $(y,z)$ and $(y',z)$ in $\C_i^{(p,r)}$ and $\C_j^{(p',r')}$ respectively. Without loss of generality, assume that $z$ is the first vertex in $\C_j^{(p',r')}$ and the last vertex in $\C_i^{(p,r)}$, and assume $(y,z)$ and $(y',z)$ occupy the same cone $C_k^z$. For both $(y,z)$ and $(y',z)$ to be added in $AddCanonical(\cdot, \cdot)$ \ref{aux-edge}, it must be that $k = i-2 = j+2$. Without loss of generality let $i=0$,$k=4$ and $j=2$, and $z$ is in $C_0^p$ and $C_2^{p'}$. 
	
	We know that both $\triangle(pyz)$ and $\triangle(p'y'z)$ are triangles in $DT(P)$. Since $(y,z) \in C_1^y$, and $(y,p) \in C_3^y$, there is no edge in $C_1^y$ clockwise from $(y,z)$. Symmetrically, there is no edge in $C_1^{y'}$ counter-clockwise from $(y',z)$. Thus $y$ must have a neighbour closer than $z$ in $C_1^y$ counter-clockwise from $(y,z)$, and $y'$ must have a neighbour closer than $z$ in $C_1^{y'}$ clockwise from $(y',z)$. See Figure \ref{fig-one-canonical-1}.
	
	\begin{figure}%
		\centering
		\subfloat[There must be an edge of $\EA$ between $(y,z)$ and $(y',z)$.] {{\includegraphics[page=1, width = 5.5cm]{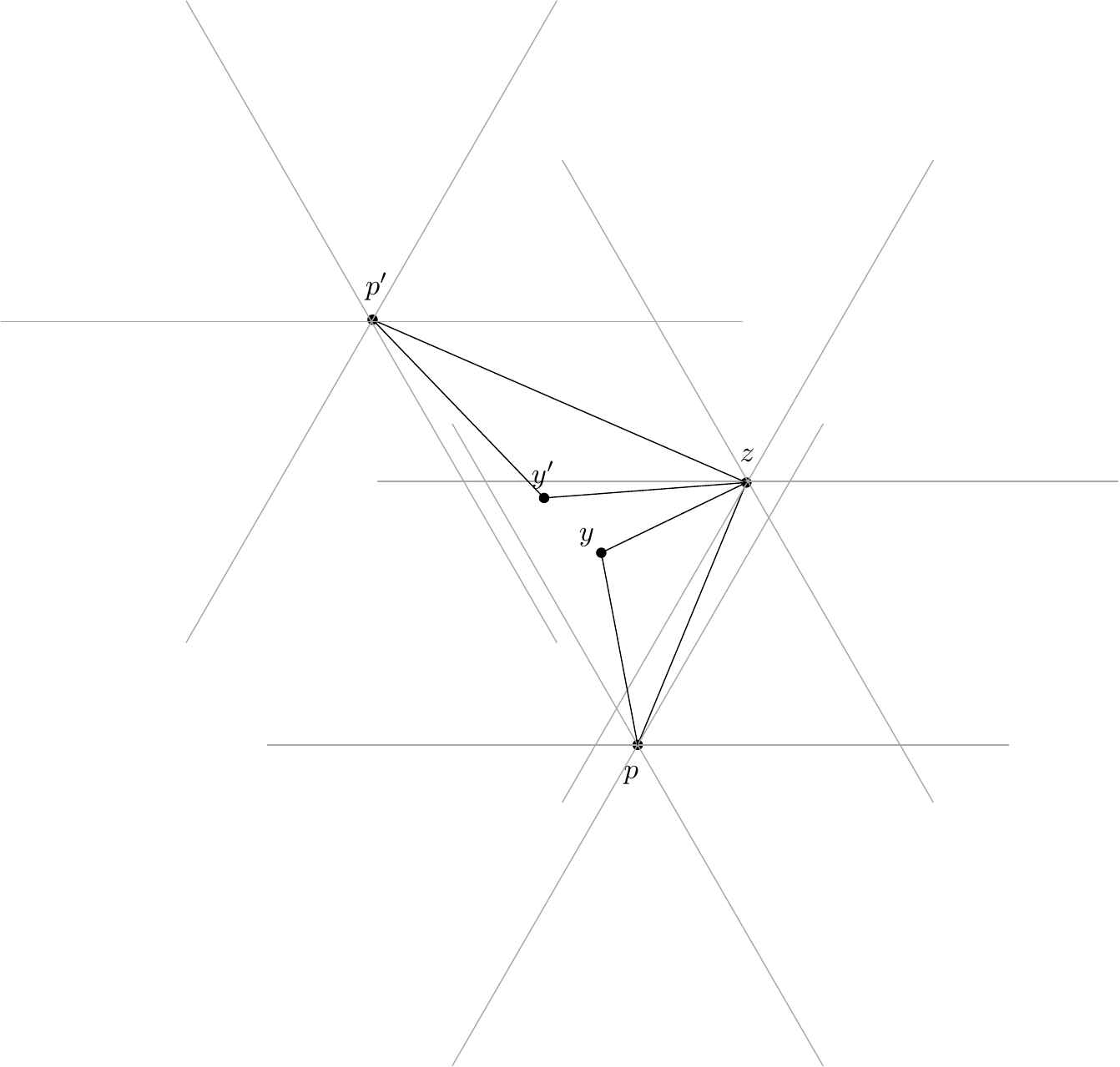} }\label{fig-one-canonical-1}}%
		\qquad
		\subfloat[If $(y,z)$ is not in $\EA$, there must be a neighbour of $y$ or $z$ in $\triangle(yxz)$.] {{\includegraphics[page=2, width = 5.5cm]{pics/one-canonical.pdf} }\label{fig-one-canonical-2}}%
		\caption{Lemma \ref{lemma-double-canonical}}
	\end{figure}
	
	We consider the shorter of $(y,z)$ and $(y',z)$, ties broken arbitrarily. Without loss of generality we will assume $[yz]<[y'z]$. We therefore know that $y' \notin C^y_1$. So there must be a vertex $t$ in $C_1^y$ that is a neighbour of $y$ and closer to $y$ than $z$ counter-clockwise from $(y,z)$. Since $z \in C_1^y$ and $y'$ is not, the counter-clockwise cone boundary of $C_1^y$ must intersect $(y',z)$ at a point which we will call $x$. Therefore $t$ must be in triangle $\triangle(xyz)$. See Figure \ref{fig-one-canonical-2}.
	
	Within $\triangle(xyz)$ we take the closest vertex to $z$ and call it $u$. $(u,z)$ must be an edge in $DT(P)$, and $C_1^u$ is bounded on one side by $(y,z)$, and bounded on the other side by $(y',z)$, and thus $z$ is the closest point to $u$ in $C_1^u$. Which means that $(u,z)$ would have been added to $\EA$ in $AddIncident()$, which means that $C_4^z$ has an edge $(u,z)\in \EA$. Since there is an edge of $\EA$ in $C_4^z$, neither $(y,z)$ nor $(y',z)$ would have been added to $\EB$ in calls to $AddCanonical(\cdot, \cdot)$ \ref{aux-edge}, and neither would be charged to $C_4^z$.
	
\end{proof}

This leads to the following corollary:

\begin{corollary}
	Assume an edge $(y,z)$ is added to $\EB$ in $AddCanonical(p,r)$ \ref{aux-edge}, and charged to a boundary cone $C_4^z$. Then of all the edges in $D8(P)$, only $(y,z)$ is charged to $C_4^z$.
\end{corollary}


The shared triangle is the only scenario where the internal cones of two separate \ncs are adjacent on the same vertex. 

Consider a shared triangle $\triangle(pp's)$ with base $(p,p')$, and assume that $p'$ is in $C_0^p$, $(p,r)\in\EA$ is in $C_0^p$, and $(p',r')\in \EA$ is in $C_3^{p'}$. Vertex $s$ has adjacent cones inside $\C_0^{(p,r)}$ and $\C_3^{(p',r')}$. We prove a limit on the number of canonical edges of $p$ and $p'$ that were added to $\EB$ and charged to cones of $s$ inside $\C_0^{(p,r)}$ and $\C_3^{(p',r')}$.

\begin{lemma}\label{charging-shared-triangles2}
	If $(s,p')$ was added to $\EB$ and charged to the empty cone of $s$ inside $\C_0^{(p,r)}$, then $(s,p)$ will not be charged to the empty cone of $s$ inside $\C_3^{(p',r')}$.
\end{lemma}

\begin{figure}
	\centering
	\subfloat[Only $(p,s)$ can be added to $\EB$ in \ref{aux-path} with apex $q$.]{	\includegraphics[width = 4.5cm]{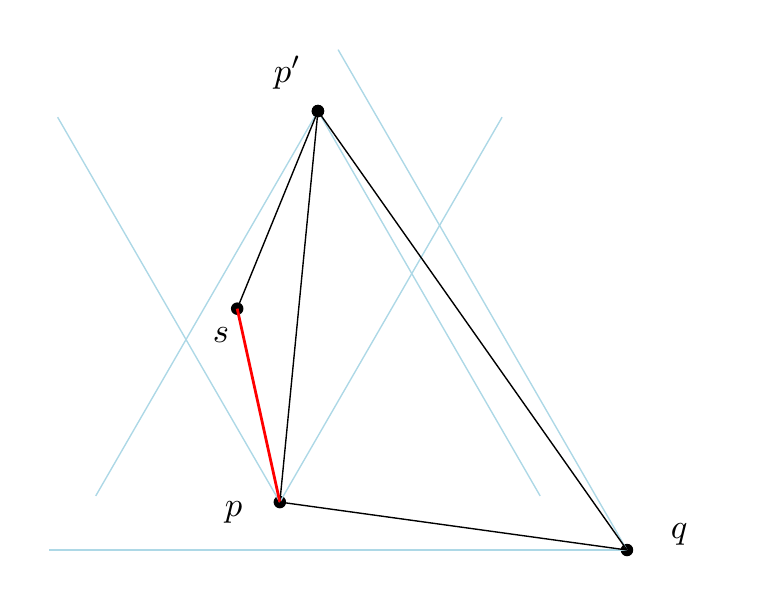}\label{fig-sharedtriangle3}}
	\qquad
	\subfloat[Only $(p',s)$ can be added to $\EB$ in \ref{aux-path} with apex $q$.]{	\includegraphics[width = 4.5cm]{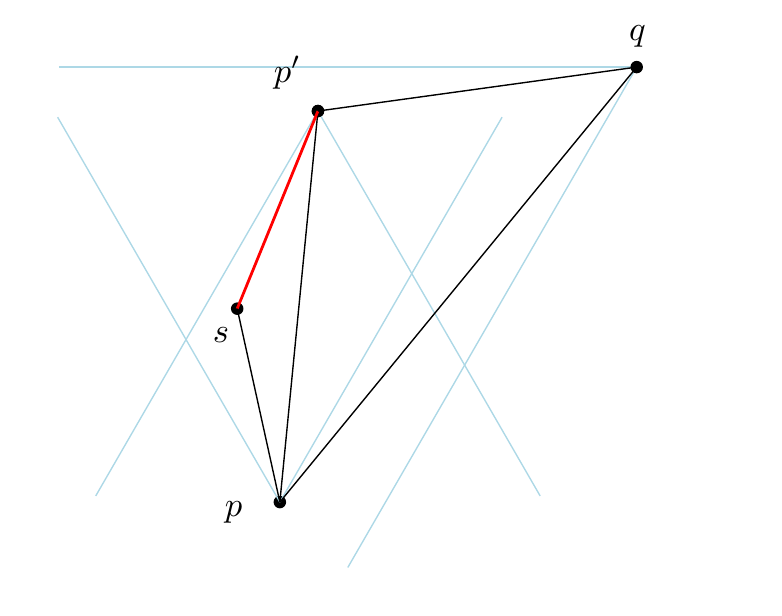}\label{fig-sharedtriangle3.1}}
	\caption{The limit on edges added to $\EB$ in a shared triangle.}
\end{figure}

\begin{proof}
	Assume that $(p',s)$ was added to $\EB$ by a call to $AddCanonical(p,r)$. That implies that $(p',s)$ is not the first or last edge of $\C_i^{(p,r)}$. Thus we know by Lemma \ref{lemma-two-shared-triangles} that $(p,s)$ must be the last edge in $\C_3^{(p',r')}$, which implies that it is not added by $AddCanonical(p',r')$.
	
	Otherwise assume that $(p,p')$ is a canonical edge of $q$, and $(p,s)$ was added to $\EB$ on a call to $AddCanonical(q,\cdot)$ in \ref{aux-path}. This implies that $(p,p')$ is in $C_5^q$. See Figure \ref{fig-sharedtriangle3}. There are two possible ways to add $(p',s)$ so that it is charged to the cone of $s$ inside $\C_3^{(p',r')}$. We show that neither occurs:
	
	\begin{enumerate}
		\item $AddCanonical(q,\cdot)$ adds $(p',s)$ to $\EB$ in \ref{aux-path}. This implies that $(p,p')$ is in $C_4^q$. See Figure \ref{fig-sharedtriangle3.1}. However, $(p,s)$ was added in \ref{aux-path}, which means that $(p,p')$ is in $C_5^q$, which is a contradiction. Thus both edges cannot be added by calls to $AddCanonical(q,\cdot)$, \ref{aux-path}.
		
		\item $AddCanonical(p,r)$ adds $(p',s)$ to $\EB$. The shared neighbour $q$ of $p$ and $p'$ is not in $C_0^p$, and thus $(p',s)$ is the last canonical edge in $\C_0^{(p,r)}$. Thus $(p',s)$ is not added to $\EB$ by a call to $AddCanonical(p,r)$ (by omission, \ref{step-duos}).
	\end{enumerate}
\end{proof}

\begin{corollary}\label{at-most-three}
	If the empty cone of $s$ inside $\C_0^{(p,r)}$ is charged twice by edges of $\C_0^{(p,r)}$, then the empty cone of $s$ inside $\C_3^{(p',r')}$ is charged at most once by edges of $\C_3^{(p',r')}$.
\end{corollary}


\begin{lemma} \label{unique-cones}
	All cones charged in the charging scheme are unique to their referenced \canp.
\end{lemma}

\begin{proof}
	We note that all the edges added here are from a \canp, thus all the charges are to vertices of a \canp, and thus must be to an inner vertex, an anchor, or an end vertex. By Lemma \ref{lemma-empty-cone} each inner vertex has an empty cone unique to its \can. By Lemma \ref{lemma-empty-cone-end}, if the end vertex has an empty cone it is unique to its \can, and by Lemma \ref{lemma-empty-cone-anchor}, the two possible empty cones on an anchor are unique to it \can.
	
	If an edge is added to $\EB$ in  $AddCanonical(p, r)$ \ref{aux-edge} it is charged to the boundary cone that it occupies. By Lemma \ref{lemma-double-canonical} it is the only edge charged to that cone, thus we consider it unique to its \can.  
\end{proof}

\begin{lemma} \label{single-charged}
	Cones of an end vertex or anchor of a \can are charged at most once by edges of $\EB$.
\end{lemma}

\begin{proof}
	Lemma \ref{unique-cones} proves that all cones charged in the charging scheme are unique (to the referenced \can). Since cones of end vertices or anchors are charged at most once in the charging scheme, this implies the lemma. 
\end{proof}

\begin{lemma} \label{double-charged}
	Cones on an inner vertex of a \can are charged at most twice by edges of $\EB$.
\end{lemma}

\begin{proof}
	Lemma \ref{unique-cones} proves that all cones charged in the charging scheme are unique (to the referenced \can). Since cones of inner vertices are charged at most twice in the charging scheme, this implies the lemma. \qed
\end{proof}

\begin{lemma}\label{no-edge-in-cone}
	The edges of $\EA$ and $\EB$ are never charged to the same cone. 
\end{lemma}

\begin{proof}
	The edges of $\EA$ are charged directly to the cone they occupy on each endpoint. We know from the charging scheme  that the edges of $\EB$ are charged to either empty cones, or to a cone that does not contain an edge of $\EA$. Thus the edges of $\EB$ and $\EA$ are never charged to the same cone. 
\qed \end{proof}

\begin{lemma}\label{bounding-cones}
	Consider a cone $C_i^s$ of a vertex $s$ in $D8(P)$ that is charged twice by edges of $\EB$. Then the two neighbouring cones $C_{i-1}^s$ and $C_{i+1}^s$ are charged at most once by edges of $D8(P)$. 
\end{lemma}

\begin{proof}
	Lemmas \ref{lemma-deg6}, \ref{single-charged}, \ref{double-charged}, and \ref{no-edge-in-cone} state that only a cone on an inner vertex may be double charged. 
	
	Each cone $C_{i-1}^s$ and $C_{i+1}^s$ is either an empty internal cone of $\C_i^{(p,r)}$, or a boundary cone containing a canonical edge of $\C_i^{(p,r)}$ with endpoint $s$. We will consider $C_{i+1}^s$ since the other cases are symmetric.
	
	If $C_{i+1}^s$ is an empty internal cone of $\C_i^{(p,r)}$, then it is only charged for an edge if $s$ is on a shared triangle $\triangle(pp's)$ and $s$ is not on the base. In this case $C_{i+1}^s$ is charged for at most one edge of $\EB$ by Lemma \ref{at-most-three}. 
	
	Otherwise $C_{i+1}^s$ contains a canonical edge in $\C_i^{(p,r)}$. By our charging scheme and Lemma \ref{at-most-three} we know only empty cones are double charged, and by Lemma \ref{no-edge-in-cone} no cone is charged for both an edge of $\EA$ and an edge of $\EB$. Thus $C_{i+1}^s$ is either charged for an edge of $\EA$, an edge of $\EB$, or it is not charged.
\qed \end{proof}
\begin{theorem}\label{lemma-d8}
	The maximum degree of $D8(P)$ is at most 8. 
\end{theorem}

\begin{wrapfigure}[18]{r}{6cm}
	\vspace{-0.8cm}
	\centering
	\includegraphics[width = 6cm]{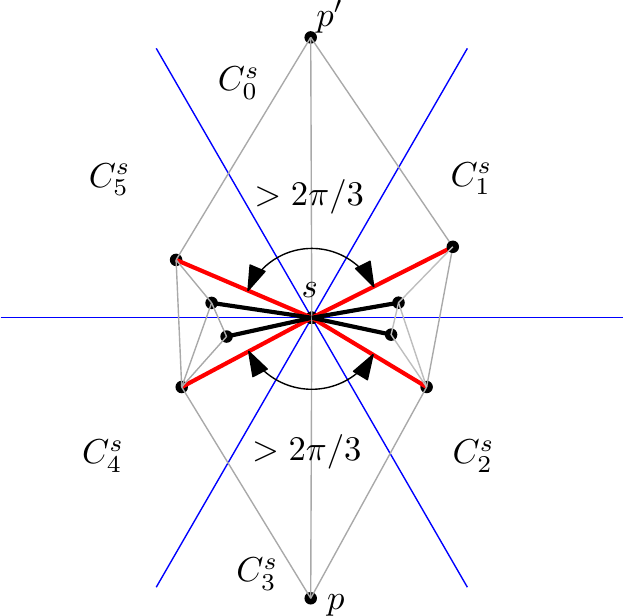}
	\caption{A degree 8 vertex in $D8(P)$. The red edges  belong to $\EB$, while the black edges belong to $\EA$.}\label{constr}
\end{wrapfigure}

\begin{proof}	
	Each edge $(p,r)$ of $\EA$ is charged once to the cone of $p$ containing $r$ and once to the cone of $r$ containing $p$. By Lemma \ref{lemma-deg6}, no cone is charged more than once by edges of $\EA$. 
	
	No edge of $\EB$ is charged to a cone that is charged by an edge of $\EA$ by Lemma \ref{no-edge-in-cone}. 
	
	By Lemma \ref{bounding-cones}, if a cone of a vertex $s$ of $D8(P)$ is charged twice, then its neighbouring cones are charged at most once. This implies that there are at most 3 double charged cones on any vertex $s$ in $D8(P)$.
	
	Assume that we have a vertex $s$ with 3 cones that have been charged twice. A cone of $s$ that is charged twice is an internal cone of some \nc $N_i^p$ by our charging argument. Thus $s$ is endpoint to two canonical edges $(q,s)$ and $(s,t)$ in $N_i^p$. Note that $\angle(qst)>2\pi/3$ by Lemma \ref{lemma-wedge-opposite-angles}, and this angle contains the cone of $s$ that is charged twice. Thus to have 3 cones charged twice, the total angle around $s$ would need to be $>2\pi$, which is impossible. Thus there are at most two double charged cones on $s$, which gives us a maximum degree of 8. See Fig. \ref{constr} for an example of a degree 8 vertex. 
	
\qed \end{proof}

\newpage
\section{D8(P) is a Spanner}\label{chap-spanner}

We will prove that $D8(P)$ is a spanner of $DT(P)$ with a spanning ratio of 
$\tspan = \ospan \approx 2.21$, thus making it a $\ospan\cdot C_{DT}$-spanner of the complete 
geometric graph, where $C_{DT}$ is the spanning ratio of the Delaunay 
triangulation. As of this writing, the current best bound of the spanning ratio of the Delaunay triangulation is 1.998\cite{xia}, which makes $D8(P)$ approximately a $4.42$-spanner of the complete graph. 

Suppose that $(p,q)$ is in $DT(P)$ but not in $D8(P)$. We will show the existence of a short path between $p$ and $q$ in $D8(P)$. If the short path from $p$ to $q$ consists of the ideal situation of an edge $(p,r)$ of $\EA$ in the same cone of $p$ as $q$, plus every canonical edge of $p$ from $r$ to $q$, then we have what we call the \emph{ideal path}. We give a spanning ratio of the \ip with respect to the \emph{canonical triangle} $T_{pq}$, which, informally, is an equilateral triangle with vertex $p$ and height $[pq]$. Notice that in our construction, when adding canonical edges to $\EB$ on an edge $(p,r)$ of $\EA$, there are times where the first or last edges of $\C_i^{(p,r)}$ are not added to $\EB$. In these cases we prove the existence of alternate paths from $p$ to $q$ that still have the same spanning ratio. Finally we prove that the spanning ratio given in terms of the canonical triangle $T_{pq}$ has an upper bound of $(1+\theta/\sin\theta)|pq|$, where $\theta = \pi/3$ is the cone angle. A canonical triangle $T_{pq}$ is the equilateral triangle with $p$ at one corner, contained in the cone of $p$ that contains $q$, and has height $[pq]$.

\subsection{Ideal Paths}\label{ideal-paths}

 We begin by defining the \ipp, and proving the spanning ratio of an \ip with respect to the graph $DT(P)$. 

\begin{definition}
	Consider an edge $(p,r)$ in $C_i^p$ in $\EA$, and the graph $\C_i^{(p,r)}$. An \emph{\ipp} is a simple path from $p$ to any vertex in $\C_i^{(p,r)}$ using the edges of $(p,r) \cup \C_i^{(p,r)}$. 
\end{definition}

  Consider an edge $(p,r)$ in $C_i^p$ in $\EA$, and the graph $\C_i^{(p,r)}$.  We will prove that the length of the \ip from $p$ to $q$ is not greater than $|pa|+\frac{\theta}{\sin\theta}|aq|$, where $a$ is the corner of the canonical triangle to the side of $(p,q)$ that has $r$, and $\theta = \pi/3$ is the cone angle. 
  
  We then use \ips to prove there exists a path with bounded spanning ratio between any two vertices $p$ and $q$ in $D8(P)$, where $(p,q)$ is an edge in $DT(P)$. We prove a bound on the length of the path from $p$ to $q$ of $\halfspan$.

  We note that the distance $\halfspan$ is with respect to the canonical triangle $T_{pq}$ rather than the Euclidean distance $|pq|$. To finish the proof we show that $|pa|+\frac{\theta}{\sin\theta}|aq| \leq (1+\frac{\theta}{\sin\theta})|pq|$.

To bound the length of \ips, we first show that a \can forms a path. Then we prove the bound.

We begin with a couple of well-known geometric lemmas. The first is an observation regarding the relative lengths of convex paths, when one resides inside the other. 

\begin{lemma}\label{lemma-convex-length}
	If a convex body $C$ is contained within another convex body $C'$, then the perimeter of $C'$ is longer than $C$.\cite{euclidean}, page 42. 
\end{lemma}

The next lemma is a well known result traditionally called ``The Inscribed Angle Theorem''.

\begin{lemma}\label{lemma-inscribed-angle}
	Consider 3 points $p,q,s$ on the boundary of a circle $O$ with center $o$, such that $\angle(pqs) = \alpha$. Let $A$ be the arc of $O$ from $p$ to $s$ that does not go through $q$, and let $\overline{A}$ be the arc of $O$ from $p$ to $s$ through $q$. Then the angle $\angle(pos)$ facing $A$ is equal to $2\alpha$. Further, the angles $\angle(pqs)$ facing $A$ is the same for any point $p$ that is on $\overline{A}$.
\end{lemma}

That allows us to establish this result:

\begin{lemma}\label{lemma-arc-length}
	Let $O$ be a circle through points $p$ and $q$ and $r$ in clockwise order, and let $\alpha$ denote the angle $\angle (qpr)$. Then the length of the arc from $q$ to $r$ on the boundary of $D_{p,q,r}$ is $$\frac{\alpha}{\sin\alpha}|qr|$$.
\end{lemma}

\begin{figure}
	\centering
	\subfloat[The Inscribed Angle Theorem]{\includegraphics[width = 4cm]{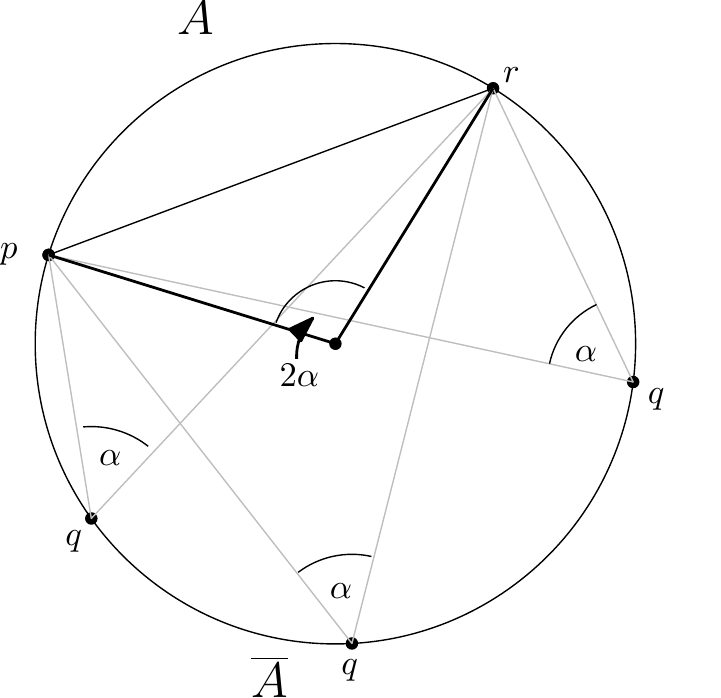}}
	\qquad
	\subfloat[Length of arc between $r$ and $q$.]{
		\includegraphics[width = 4cm]{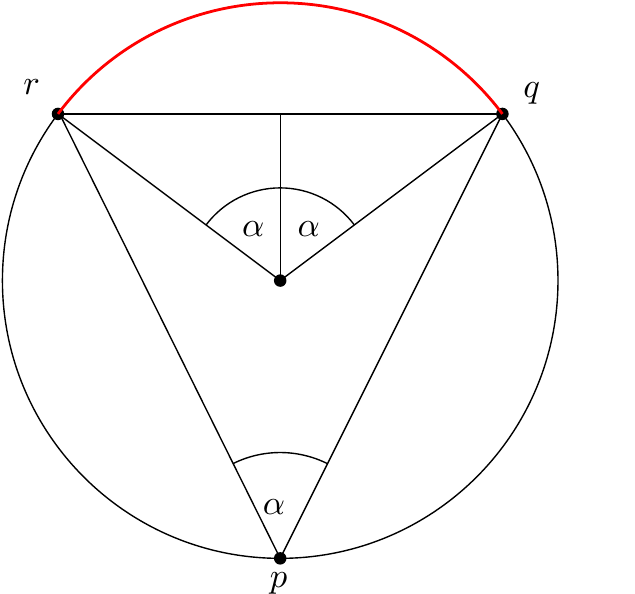}\label{arc-length}}
	\caption{Relating arc length to angle.}
\end{figure}

\begin{proof}
	From the center point of $O$, the angle between $q$ and $r$ is $2\alpha$ by Lemma \ref{lemma-inscribed-angle}. Thus the arc length between $q$ and $r$ is $2\alpha R$, where $R$ is the radius of $O$. Also, $|qr| = 2 \sin\alpha R$, which means $R = \frac{|qr|}{2\sin\alpha}$. Thus the arc length between $q$ and $r$ is equal to:
	\begin{align*}
	2\alpha R &= \frac{2\alpha}{2\sin\alpha}|qr|\\
	&= \frac{\alpha}{\sin\alpha}|qr|
	\end{align*}
	which completes the proof. See Figure \ref{arc-length}.
\end{proof}

We require that a \can is a path, which is proven here.

\begin{lemma}\label{canpath}
	Let $(p,r)$ be an edge in $\EA$ in the cone $C_i^p$. Then $\C_i^{(p,r)}$ forms a path.
\end{lemma}

\begin{proof}
	We prove by contradiction. Note that $\C_i^{(p,r)}$ is a collection of paths. Assume that there are at least two paths in this collection. Without loss of generality, let $i=0$. Let $(a,b)$ and $(y,z)$ be the first and last edge respectively in $\C_0^{(p,r)}$. Thus of all the vertices in $N_0^p \backslash \{p\}$ between $a$ and $z$ there exists at least one consecutive subset $T$ where for each $t_j \in T, 0\leq j < |T|, [pt_j]<[pr]$. We consider the vertex  $t_k\in T, [pt_k]\leq [pt_j]$, for all $t_j \in T, 0\leq j < |T|$. Since $[pt_k]<[pr]$, $AddIncident(L)$ examined $(p,t_k)$ before $(p,r)$. Thus when $(p,t_k)$ was examined, $C_i^p$ contained no edges of $\EA$ with endpoint $p$. Since $(p,t_k)$ was not added to $\EA$, there must have been an edge of $\EA$ with endpoint $t_k$ in $C_3^{t_k}$. However, we know $[pt_{k-1}]\leq[pt_k]$ and $[pt_{k+1}]\leq [pt_k]$ (whether or not $t_{k-1}$ and $t_{k+1}$ are in $T$). Thus neither $(t_k,t_{k-1})$ nor $(t_k,t_{k+1})$ can be in $C_3^{t_k}$. Since $\triangle(pt_kt_{k-1})$ and $\triangle(pt_kt_{k+1})$ are triangles in $DT(P)$, the only edge with endpoint $t_k$ in $C_3^{t_k}$ is $(p,t_k)$. This means that $(p,t_k)$ would have been added to $\EA$ instead of $(p,r)$, which is a contradiction.
	\qed \end{proof}

\begin{lemma} \label{no-points}
	Consider the \rn $N_p^{(r,q)}$ in $DT(P)$ in the cone $C_i^p$. Let $O_{p,r,q}$ be the circle through the points $p$, $q$, and $r$. Then there are no points of $P$ in $O_{p,r,q}$ to the side of $(p,r)$ that does not contain $q$. Likewise there are no points of $P$ in $O_{p,r,q}$ to the side of $(p,q)$ that does not contain $r$. 
\end{lemma}

\begin{figure}
	\centering
	\subfloat[If there is a point $t$, then $(p,r)$ is not an edge in $DT(P)$.]{\includegraphics[width = 4cm, page =3]{pics/new-keil.pdf}\label{new-keil2}}
	\qquad
	\subfloat[Circle $O_{r,a,q}$ contains $u$.]{
		\includegraphics[width = 3cm]{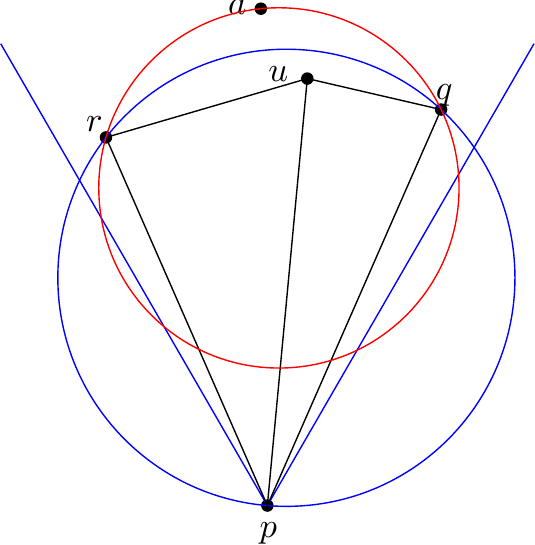}\label{intermediate1}}
	\qquad
	\subfloat[A circle through $a$ and $p$ contains $u$ or $v$.]{\includegraphics[width = 3cm, page = 2]{pics/intermediate.pdf}\label{intermediate2}}
	\caption{Locations of $a$.}
\end{figure}

\begin{proof}
	Since the cases are symmetric, we prove that there are no points of $P$ in the region $R$ of $O_{p,r,q}$ to the side of $(p,r)$ that does not contain $q$. We prove by contradiction. Thus assume there is a point $t$ in $R$. Then the circle $O_{p,t,r}$ contains $q$ and the circle $O_{p,r,q}$ contains $t$, thus there is no circle through $p$ and $r$ that is empty of points of $P$. Thus $(p,r)$ cannot be a Delaunay edge, which is a contradiction to our definition of \rnp. See Fig. \ref{new-keil2}.
	\qed \end{proof}

\begin{lemma}\label{intermediate}
	Consider the \rn $N_p^{(r,q)}$ in cone $C_i^p$. Let $rq$ be the directed line from $r$ to $q$, and assume there are no neighbours of $p$ in $N_p^{(r,q)}$ right of $rq$. If $(r,q)$ is not an edge in $N_p^{(r,q)}$, then there is a vertex $a\in N_p^{(r,q)}$ such that the circle $O_{r,a,q}$ is empty of vertices of $P$ left of $rq$. 
\end{lemma}

\begin{proof}
	
	We prove by contradiction, thus assume that we have found a vertex $a$ left of $rq$ such that $O_{r,a,q}$ is empty of vertices of $P$ left of $rq$, and $a$ is not in $N_p^{(r,q)}$. Note vertex $a$ must exist, otherwise $(r,q)$ is on the convex hull and thus in $N_p^{(r,q)}$. Since the region of $N_p^{(r,q)}$ is empty of vertices of $P$, $a$ must be outside of $N_p^{(r,q)}$.
	
	We look at two cases:
	
	\begin{enumerate}
		\item $a$ is outside of $O_{p,r,q}$: Since $(r,q)$ is not in $DT(P)$, there is at least one vertex $u$ in $N_p^{(r,q)}\backslash\{p,r,q\}$. By Lemma \ref{lemma-inside-edge-in-disk} and our initial assumption that $N_p^{(r,q)}$ contains no neighbours of $p$ to the right of $rq$, $u$ must be in $O_{p,r,q}$ to the left of $rq$. Since $a$ is outside of $O_{p,r,q}$, the arc of $O_{r,a,q}$ to the left of $rq$ contains the arc of $O_{p,r,q}$ to the left of $rq$. Thus $u$ is in $O_{r,a,q}$ to the left of $rq$, which is a contradiction to our selection of vertex $a$. See Fig. \ref{intermediate1}.
		
		\item $a$ is inside $O_{p,r,q}$: Since $\angle(rpq) <\pi/3$ (since it is in a cone), and $a$ is inside $O_{p,r,q}$, $a$ must be positioned radially between two consecutive edges with endpoint $p$ in $N_p^{(r,q)}$. Call these edges $(p,u)$, and $(p,v)$. Note that $\triangle(puv)$ is a triangle in $DT(P)$, and thus the circle $O_{p,u,v}$ does not contain $a$ by the empty circle property of the Delaunay triangulation. This implies that, since $p$,$u$,$a$, and $v$ form a convex quadrilateral with $p$ and $a$ across the diagonal, any circle through $p$ and $a$ must contain at least one of $u$ or $v$. 
		
		Since $a$ is inside $O_{p,r,q}$, $O_{r,a,q}$ contains $p$. Thus we can draw the circle $O_1$ through $a$ and $p$ tangent to $O_{r,a,q}$. The portion of $O_1$ to the left of $rq$ is contained in $O_{r,a,q}$, and thus does not contain any points of $P$. But any circle through $a$ and $p$ must contain at least one of $u$ and $v$, and $u$ and $v$ are to the left of $rq$, which is a contradiction. See Fig. \ref{intermediate2}
		
	\end{enumerate}
	Thus, if $(r,q)$ is not an edge in $N_p^{(r,q)}$, there is a neighbour $a$ of $p$ in $N_p^{(r,q)}$ such that $O_{r,a,q}$ is empty of vertices of $P$ left of $rq$. See Fig. \ref{intermediate2}.
	
	\qed \end{proof}

We now turn to a lemma from the paper of Bose and Keil\cite{bosekeil} that tells us the length of a path between two points in the Delaunay triangulation of a set of vertices. We provide a slightly modified and truncated version that suits our needs. The lemma of Bose and Keil does not provide an explicit construction. We apply the lemma to a \rnp, and are able to provide a construction of the path along with an upper bound on its length.

\begin{lemma}\label{rq-path}
	Consider the \rn $N_p^{(r,q)}$ in $DT(P)$ in the cone $C_i^p$. Let $\alpha = \angle(rpq)< \pi/3$. If no point of $P$ lies in the triangle $\triangle(prq)$ then there is a path from $r$ to $q$ in $DT(P)$, using canonical edges of $p$, whose length satisfies:
	$$\delta(r,q) \leq |rq|\frac{\alpha}{\sin \alpha}$$ 
\end{lemma}

\begin{figure}
	\centering
	\includegraphics[width = 5.5cm]{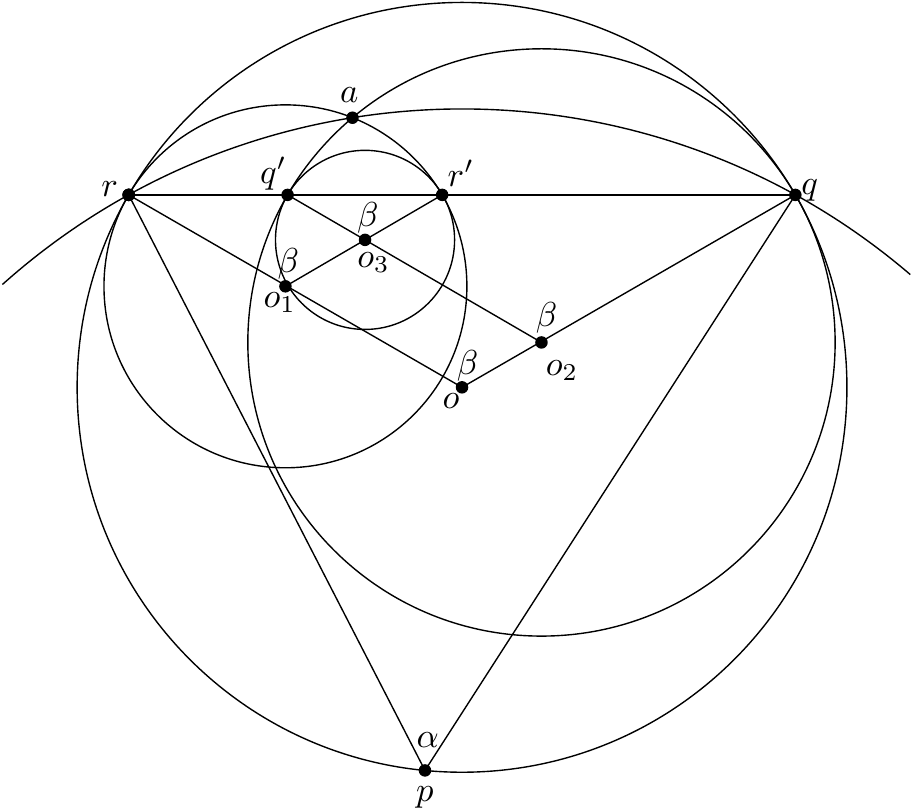}\caption{Lemma \ref{rq-path}.}\label{keils-lemma}
\end{figure}

\begin{proof}
	
	Let $o$ be the center of $O_{p,r,q}$, and let  $\beta = \angle(roq) = 2\alpha$. 
	
	Lemma \ref{no-points} and the assumption that no vertices of $P$ lie in the triangle $\triangle(prq)$ imply that there are no vertices of $P$ in $O_{p,r,q}$ to the right of directed line segment $rq$.
	
	We proceed by induction on the number of vertices in $N_p^{(r,q)}$. If there are only 3 vertices in $N_p^{(r,q)}$, then $(r,q)$ is an edge in $DT(P)$, and the path from $r$ to $q$ has length $|rq|<|rq|\frac{\alpha}{\sin \alpha}$ and we are done. 
	
	Now assume that the inductive hypothesis holds for all \rns with fewer vertices than $N_p^{(r,q)}$. Assume $N_p^{(r,q)}$ has more than 3 vertices, otherwise we are done by the same argument as above.
	
	Lemma \ref{no-points} tells us that there is a vertex $a$ in $N_p^{(r,q)}$ where $O_{r,a,q}$ is empty of vertices of $P$ left of $rq$. 	
	
	Let $O_1$ be the circle through $r$ and $a$ with center $o_1$ on the line segment $(o,r)$. Let $O_2$ be the circle through $a$ and $q$ whose center $o_2$ lies on the line segment $(o,q)$. Let $\alpha_1 = \angle(ro_1a)$ and let $\alpha_2 = \angle(ao_2q)$.  $N_p^{(r,a)}$ and $N_p^{(a,q)}$ have fewer vertices than $N_p^{(r,q)}$, and $O_1$ is empty of vertices of $P$ to the right of directed segment $ra$, and $O_2$ is empty of vertices of $P$ to the right of directed line segment $aq$. Thus by the inductive hypothesis:
	
	\begin{align*}
	\delta(r,q)&=\delta(r,a) +\delta(a,q)\\
	&= |ra|\frac{\alpha_1}{\sin \alpha_1} +|aq|\frac{\alpha_2}{\sin \alpha_2}
	\end{align*}

	Let $r'\neq r$ be the intersection of $O_1$ and $rq$, and let $q'\neq q$ be the intersection of $O_2$ and $rq$. Since $\beta < \pi$, $O_1$ and $O_2$ overlap. Let $O_3$ be the circle through $q'$ and $r'$ with center $o_3$ on the intersection of the line segment between $o_1$ and $r'$ and the line segment between $o_2$ and $q'$. See Fig. \ref{keils-lemma}.	
	
	Triangles $\triangle(roq)$, $\triangle(ro_1r')$, $\triangle(q'o_2q)$, and $\triangle(q'o_3r')$ are all similar isosceles triangles. Thus by Lemmas \ref{lemma-inscribed-angle} and \ref{lemma-arc-length} the length of the arc of $O_1$ left of $rq$ is $|rr'|\frac{\alpha}{\sin \alpha}$, the length of the arc of $O_2$ left of $rq$ is $|q'q|\frac{\alpha}{\sin \alpha}$, and the length of the arc of $O_3$ left of $rq$ is $|q'r'|\frac{\alpha}{\sin \alpha}$.
	
	Note that $O_3$ is completely contained in the intersections of $O_1$ and $O_2$. Let $A_1$ be the arc of $O_1$ left of $rq$ from $a$ to $r'$, and let $A_2$ be the arc of $O_2$ left of $rq$ from $a$ to $q'$. Note that $A_1\cap A_2$ is a convex shape from $q'$ to $r'$ that contains the arc of $O_3$ left of $rq$. Thus $|A_1\cap A_2|\geq|q'r'|\frac{\alpha}{\sin \alpha}$ by convexity (Lemma \ref{lemma-convex-length}).
	
	We observe that:
	
	\begin{align*}
	\delta(r,q)&=\delta(r,a) +\delta(a,q)\\
	&= |ra|\frac{\alpha_1}{\sin \alpha_1} +|aq|\frac{\alpha_2}{\sin \alpha_2}\\
	&= |rr'|\frac{\alpha}{\sin \alpha} + |q'q|\frac{\alpha}{\sin \alpha} - |A_1\cap A_2|\\
	&\leq |rr'|\frac{\alpha}{\sin \alpha} + |q'q|\frac{\alpha}{\sin \alpha} - |q'r'|\frac{\alpha}{\sin \alpha}\\
	&=|rq|\frac{\alpha}{\sin \alpha}
	\end{align*}
	
	as required.

	\qed \end{proof}

\begin{lemma}\label{path}
	The path $\delta(r,q)\leq |rr_q|+|r_qq|\frac{\theta}{\sin\theta}$
\end{lemma}

\begin{proof}
	By convexity.
	\qed \end{proof}
Now we prove the following:

\begin{lemma} \label{inward-path3} 
	Consider the \rn $N_p^{(r,q)}$ and without loss of generality let $N_p^{(r,q)}$ be in $C_0^p$. Let $\alpha = \angle(rpq)$. Let $r_q\neq p$ be the point where the line through $p$ and $r$ intersects the canonical triangle $T_{pq}$. Let $q_r\neq p$ be the point where the edge $(p,q)$ intersects $T_{pr}$. If $[pr]$ is the shortest edge of all edges in $N_p^{(r,q)}$ with endpoint $p$, then the distance from $r$ to $q$ using the canonical edges of $p$ in $N_p^{(r,q)}$ is at most $\max\{|rr_q|,|q_rq|\}+|r_qq|\frac{\theta}{\sin \theta}$.
\end{lemma}

\begin{figure}%
	\centering
	\subfloat[If $|pa|\geq|pq|$, apply Lemma \ref{rq-path}.] {{	\includegraphics[page=2,width = 4cm]{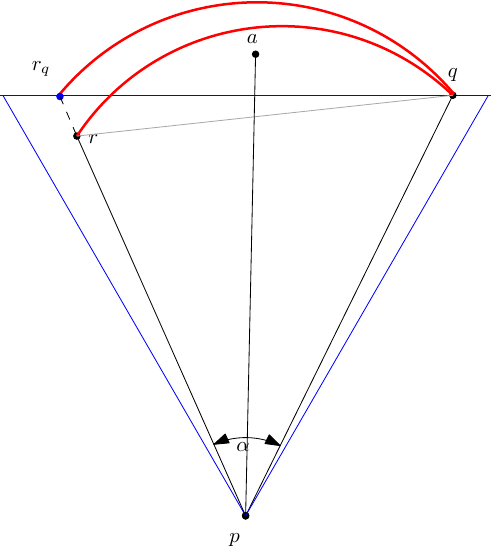} }\label{fig-inward2}}%
	\qquad
	\subfloat[If the vertex $a$ is in $\triangle(prq)$, we proceed by induction from $r$ to $a$.] {{		\includegraphics[page=4,width = 4cm]{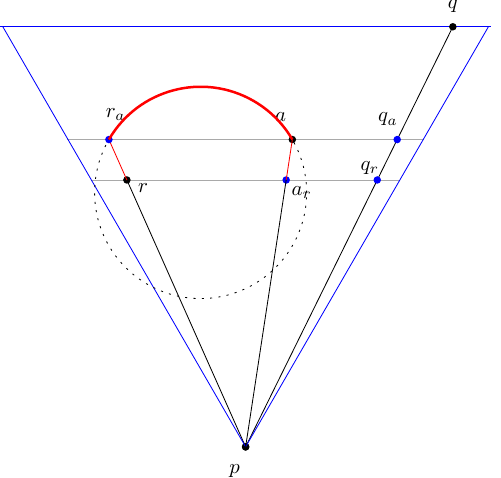} }\label{fig-inward4}}%
	\qquad
	\subfloat[Proceed by induction from $a$ to $q$.] {{		\includegraphics[page=5,width = 4cm]{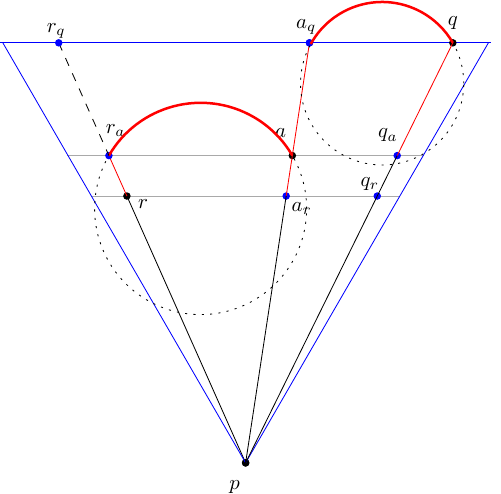} }\label{fig-inward5}}%
	\qquad
	\subfloat[$|r_aa|\frac{\theta}{\sin \theta} <|r_qa_q|\frac{\theta}{\sin \theta}$.] {{		\includegraphics[page=6,width = 4cm]{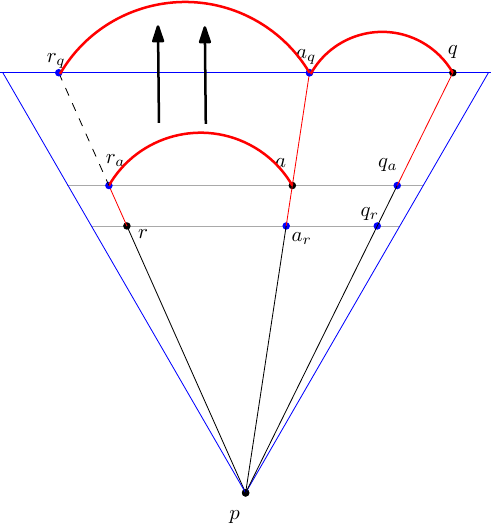} }\label{fig-inward6}}%
	\qquad
	\subfloat[$|r_qq|\frac{\theta}{\sin \theta}=|r_qa_q|\frac{\theta}{\sin \theta}+|a_qq|\frac{\theta}{\sin \theta}$.] {{\includegraphics[page=7,width = 4cm]{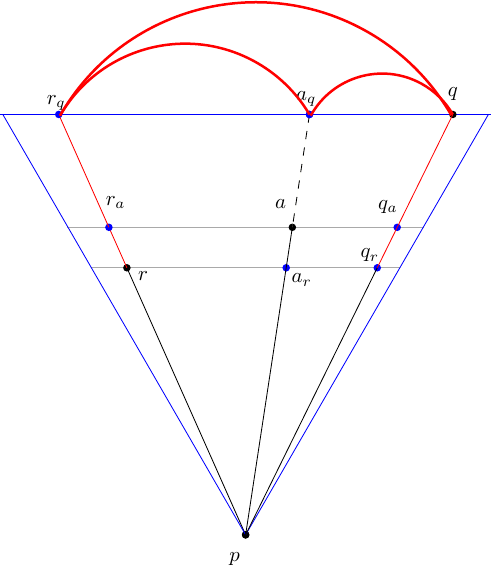} }\label{fig-inward7}}%
	\qquad
	\subfloat[Lemma \ref{lemma-path}.] {{\includegraphics[page=8,width = 4cm]{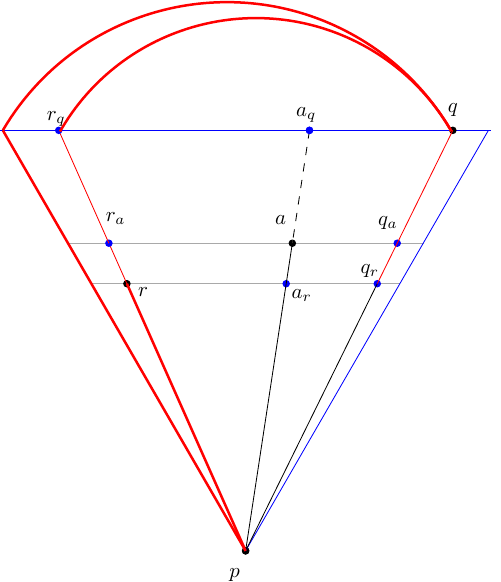} }\label{fig-inward9}}%
	\caption{Inductive path.}\label{inductive-path}
\end{figure}

\begin{proof}
	Let $\delta(r,q)$ be the length of the path between $r$ and $q$ in $N_p^{(r,q)}$. We will prove by induction on the number of canonical edges of $p$ in $N_p^{(r,q)}$. 
	
	If there is only one canonical edge of $p$ in $N_p^{(r,q)}$, then $(r,q)$ is that edge and $\delta(r,q) = |rq|\leq 	\max\{|rr_q|,|q_rq|\}+|r_qq|\frac{\alpha}{\sin \alpha}$, we are done. 
	
	Otherwise assume there is more than one canonical edge of $p$ in $N_p^{(r,q)}$. Consider the edge $(p,a)\in N_p^{(r,q)}$, such that $[pa]\leq [pt]$, for all $(p,t)\in N_p^{(r,q)}\backslash \{r,q\}$. We consider two cases:
	
	\begin{enumerate}
		\item If $[pa]>[pq]$, then $[pr]$ and $[pq]$ are the shortest edges in $N_p^{(r,q)}$, which implies that there are no points in $\triangle(prq)$. Thus from Lemma \ref{path}, the length of the path from $r$ to $q$ is at most $|rq|\frac{\theta}{\sin \theta}$. We have
		
		\begin{align*}
		\delta(r,q)&\leq |rq|\frac{\alpha}{\sin \alpha}\\
		&\leq |rr_q|+|r_qq|\frac{\alpha}{\sin \alpha}
		\end{align*}
		
		by convexity (Lemma \ref{lemma-convex-length}).  $\frac{\alpha}{\sin \alpha}$ is increasing in $\alpha$, thus $\frac{\alpha}{\sin \alpha}\leq \frac{\theta}{\sin \theta}$. Thus
		
		\begin{align*}
		\delta(r,q)&\leq|rr_q|+|r_qq|\frac{\alpha}{\sin \alpha}\\
		&\leq \max \{|rr_q|, |q_rq|\}+|r_qq| \frac{\theta}{\sin \theta}
		\end{align*} 
		
		which satisfies the inductive hypothesis. See Fig. \ref{fig-inward2}

		\item $[pa]<[pq]$. Since $[pr]\leq[pa]$ we can apply the inductive hypothesis on $N_p^{(r,a)}$. Let $r_a$ be the point where the line through $p$ and $r$ intersects the horizontal line through $a$, and let $a_r$ be the point where the line through $p$ and $a$ intersects the horizontal line through $r$. See Fig. \ref{fig-inward4}. Then by the inductive hypothesis:
		
		\begin{align*}
		\delta(r,a) = \max\{|rr_a|,|a_ra|\}+ |r_aa|\frac{\theta}{\sin \theta}
		\end{align*}

		Since $[pa]\leq[pq]$ we can apply the inductive hypothesis on $N_p^{(a,q)}$. Let $a_q \neq p$ be the point where the line through $a$ and $p$ exits $T_{pq}$, and let $q_a\neq p$ be the point where $(p,q)$ intersects $T_{pa}$, and let $\alpha_2 = \angle(apq)$. See Fig. \ref{fig-inward5}. Then by the inductive hypothesis:
		
		\begin{align*}
		\delta(a,q) = \max\{|aa_q|,|q_aq|\}+ |a_qq|\frac{\theta}{\sin \theta}
		\end{align*}

		Note that $|pa_q| \leq \max\{|pr_q|,|pq|\}$. Thus:
		
		\begin{align*}
		\delta(r,q) &= \max\{|rr_a|,|a_ra|\}+\max\{|aa_q|,|q_aq|\}+ |r_aa|\frac{\theta}{\sin \theta}+|a_qq|\frac{\theta}{\sin \theta}\\
		&\leq \max\{|rr_q|,|q_rq|\}+|r_aa|\frac{\theta}{\sin \theta}+|a_qq|\frac{\theta}{\sin \theta}\\
		&\leq\max\{|rr_q|,|q_rq|\}+|r_qa_q|\frac{\theta}{\sin \theta}+|a_qq|\frac{\theta}{\sin \theta}\\
		&\leq\max\{|rr_q|,|q_rq|\}+|r_qq|\frac{\theta}{\sin \theta}
		\end{align*}
		
		as required.
	\end{enumerate} 
	See Fig. \ref{inductive-path}.
	\qed \end{proof}

Using Lemma \ref{inward-path3} we can prove the main lemma of this section:
\begin{lemma}\label{lemma-path}

Consider the edge $(p,r)$ in $\EA$, located in $\C_i^p$, and the associated \can $\C_i^{(p,r)}$. Without loss of generality, assume that $i=0$. The length of the \ip from $p$ to any vertex $q$ in $\C_0^{(p,r)}$ satisfies $\delta(p,q) \leq |pa|+\frac{\theta}{\sin\theta}|aq|$, where $a$ is the corner of $T_{pq}$ such that $r \in \triangle(pqa)$, and $\theta = \pi/3$ is the angle of the cones.

\end{lemma}

\begin{proof}
	 (Refer to Fig. \ref{fig-inward9}.)
	By Lemma \ref{inward-path3} the path from $r$ to $q$ is no greater than $\max\{|rr_q|,|q_rq|\}+|r_qq|\frac{\theta}{\sin\theta}$. 
	
	Since $|pr|+ \max\{|rr_q|,|q_rq|\} \leq |pa| $ and $|aq|\geq|r_qq|$ we have 
	
	\begin{align*}
	\delta(p,q) &\leq |pr|+\max\{|rr_q|,|q_rq|\}+|r_qq|\frac{\theta}{\sin\theta}\\
		&\leq |pa|+|aq|\frac{\theta}{\sin\theta}.
	\end{align*} \qed
\end{proof}

\subsection{Paths in D8(P)}

A path in $D8(P)$ that approximates an edge $(p,q)$ of $DT(P)$ can take several forms. It may consist of the edge $(p,q)$, or it may be an \ip from $p$ to $q$, it may be the concatenation of two \ips from $p$ to $q$, or some combination of the above. We prove that $\delta(p,q)$, the length of the path in $D8(P)$ that approximates edge $(p,q) \in DT(P)$, is not longer than $\max\{|pa|+\func|aq|, |pb|+\func|bq|\}$. Points $a$ and $b$ are the top left and right corners of canonical triangle $T_{pq}$ respectively.

At this point our spanning ratio is with respect to $T_{pq}$. We then prove that $D8(P)$ is a spanner with respect to the Euclidean distance $|pq|$. 

We consider an edge $(p,q) \in DT(P)$. If $(p,q) \in D8(P)$ then the length of the path from $p$ to $q$ in  $D8(P)$ is $|pq| \leq \fullspan{|pq|}$, as required. 

Thus we assume $(p,q) \notin D8(P)$. Without loss of generality we assume $q$ is in $C_0^p$. Since $(p,q)\notin D8(P)$, there is an edge $(p,r)$ of $\EA$ in $\C_0^p$ or $(q,u)$ in $\C_3^q$ (or both), where $[pr]\leq[pq]$ and $[qu]\leq[pq]$. Otherwise $(p,q)$ would have been added to $\EA$ in $AddIncident(L)$. Without loss of generality we shall assume there is the edge $(p,r) \in \EA, [pr]\leq[pq]$, and that $(p,q)$ is clockwise from $(p,r)$ around $p$.

Let $s$ be the vertex such that $s$ is a neighbour of $q$ in $N_p^{(p,r)}$ and $s\neq p$ (but possibly $s=r$). Let $a$ be the upper left corner of $T_{pq}$, and $b$ be the upper right corner. Let $\alpha = \angle(rpq)$ and $\theta = \pi/3$ be the angle of the cones. 

\begin{lemma}\label{lemma-can-path}
	Recall that $(p,r)\in \EA$, where $r \in C_i^p$. Then there is an \ip from $p$ to any vertex $q$ in $\C_i^{(p,r)}$, where $q$ is not an end vertex of $\C_i^p$.
\end{lemma}

\begin{proof}
	In the algorithm $AddCanonical(p,r)$, we add every canonical edge of $p$ in $\C_i^{(p,r)}$ that is not the first or last edge. By Lemma \ref{canpath}, the edges of $\C_i^{(p,r)}$ form a path. Thus there is the ideal path from $p$ to any vertex $q$ in $\C_i^{(p,r)}$ that is not the first or last vertex. 
\qed \end{proof}

The next lemmas prove that, for a vertex $z$ that is the first or the last vertex of $\C_i^p$, the edge in $\C_i^p$ with endpoint $z$ cannot be in $C_i^z$. 

\begin{lemma}\label{lemma-circle-exit}
	Let $r$ and $q$ be two consecutive neighbours of $p$, in an arbitrary cone $C_i^p$. Without loss of generality, let $(p,q)$ be clockwise from $(p,r)$ in the cone $C_i^p$. If $q$ is in $C_i^r$, then all edges with endpoint $p$ in $C_i^p$ that appear after $(p,q)$ in clockwise order are longer than $[pq]$. 
\end{lemma}

\begin{figure}%
	\centering
	\subfloat[Any edge with endpoint $p$ in $N_i^p$ clockwise from $(p,q)$ must be longer than $(p,q)$.] {{				\includegraphics[width = 5.5cm]{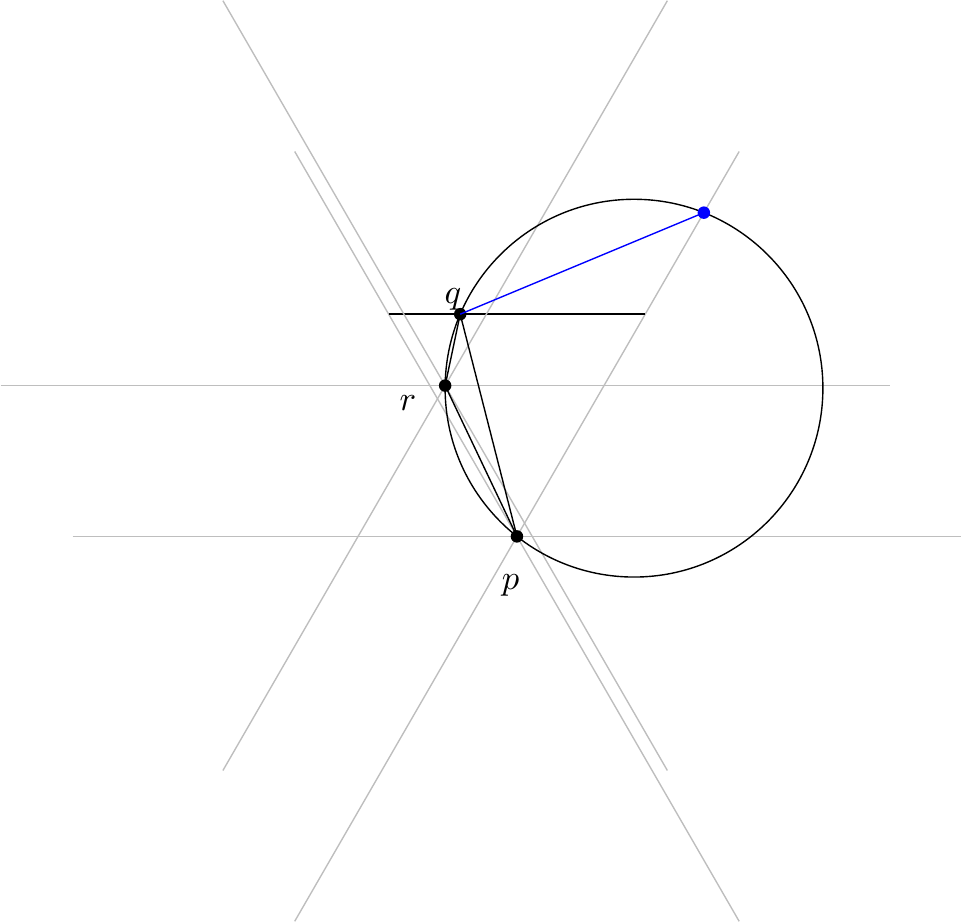} }\label{no-z}}%
	\qquad
	\subfloat[$(y,z)$ cannot be in $C_i^z$.] {{		\includegraphics[width = 4.5cm,page=1]{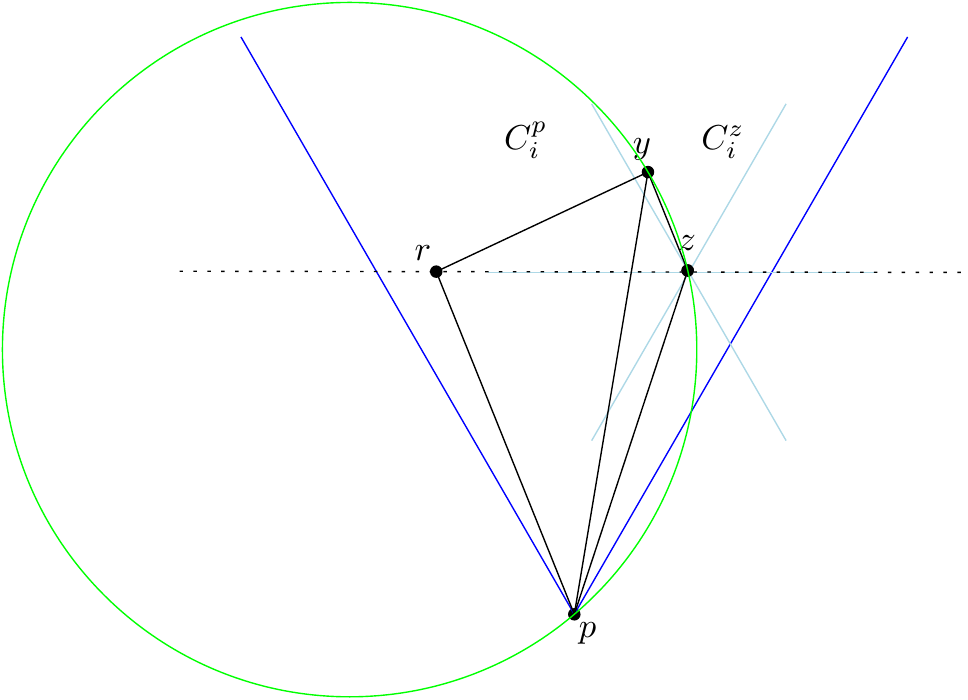} }\label{no-z2}}%
	\subfloat[If the final edge of $\C_0^p$ is in $C_0^y$, we do \emph{not} add it to $\EB$.]{{\includegraphics[page=6, width = 4cm]{pics/bisector-construction.pdf}} \label{fig-handle-last}}
	\caption{The edge in $\C_i^p$ with endpoint $z$ cannot be in $C_i^z$.}
\end{figure}

\begin{proof}	
	By Lemma \ref{lemma-wedge-opposite-angles}, any edge $(p,t)$ clockwise from $(p,q)$ in $C_i^p$ is such that the angle $\angle(rqt)>2\pi/3$. Since $(r,q)$ is in $C_i^r$, it is at an angle of at least $\pi/3$ from the positive $x$-axis. Since $\angle(rqt)>2\pi/3$, the edge $(r,t)$ must be at an angle $>0$ with respect to the positive $x$-axis. Thus $[pt]>[pq]$, for all $(p,t)$ clockwise from $(p,q)$ in $C_i^p$. See Figure \ref{no-z}. 
\end{proof}

\begin{lemma} \label{no-zero}
	Let $z$ be the first or last vertex of $\C_i^{(p,r)}$, and assume that $(p,z)$ is not in $\EA$. Let $(y,z)$ be the last edge in $\C_i^{(p,r)}$. Then $(y,z)$ is not in $C_i^z$.
\end{lemma}

\begin{proof}
	We assume that $(y,z) \in C_i^z$, and prove by contradiction. By Lemma \ref{lemma-circle-exit}, if $(y,z)$ is in $C_i^z$, then $(p,y)$ is the shortest of all edges in $C_i^p$ with endpoint $p$ counter-clockwise from $(p,y)$.
	
	Let $(p,r)$ be an edge in $\EA$, where $r \in \C_i^{(p,r)}$. Then $(p,r)$ is at least as short as all edges in $DT(P)$ from $p$ to a vertex in $\C_i^{(p,r)}$. But that is a contradiction to $(p,y)$ (and by extension $(p,z)$) being the shortest. See Fig. \ref{no-z2}
	\qed \end{proof}

Let $(p,r)$ be an edge in $\EA$ is the graph $D8(P)$. Without loss of generality, assume that $r$ is in $C_0^p$. By Lemma \ref{lemma-can-path}, there is the \ip from $p$ to any vertex in $\C_0^{(p,r)}$ that is not the first or last vertex. We now turn our attention to the first or last vertex in $\C_0^{(p,r)}$. Because the cases are symmetric, we focus on the last vertex, which we designate $z$. If $z=r$, the path from $p$ to $z$ is trivial, thus we assume $z\neq r$. Let $y$ be the neighbour of $z$ in $\C_0^{(p,r)}$. By Lemma \ref{no-zero}, $(y,z)$ cannot be in $C_0^z$. Thus $(y,z)$ can be in $C_5^z$, $C_4^z$, or $C_3^z$.

\begin{enumerate}[labelindent=*,
	style=multiline,
	leftmargin=*,label=Case \arabic*:, ref =Case \arabic*]
 \item \label{case3}Edge $(y,z)$ is in $C_5^z$. Then $(y,z)$ was added to $\EB$ in $AddCanonical(p,r)$, \ref{n5}, and there is an \ip from $p$ to $z$.  
 \item \label{case4}Edge $(y,z)$ is in $C_4^z$. There are three possibilities.
 
 \begin{enumerate}
 	\item \label{case6} If $(y,z)$ is an edge of $\EA$, then there is an \ip from $p$ to $z$. 
 	
 	\item \label{case7}If there is no edge in $\EA$ with endpoint $z$ in $C_4^z$, then $(y,z)$ was added to $\EB$ in $AddCanonical(p,r)$, \ref{aux-edge}, and there is an \ip from $p$ to $z$.   
 
	\item \label{case8} If there is an edge of $\EA$ in $C_4^z$ with endpoint $z$ that is not $(y,z)$, then we have added the canonical edge of $z$ in $C_4^z$ with endpoint $y$ to $\EB$ in $AddCanonical(p,r)$, \ref{aux-path}. Therefore by Lemma \ref{lemma-can-path} there is an \ip from $z$ to $y$, and also an \ip from $p$ to $y$. 
\end{enumerate}
 \item \label{case5}Edge $(y,z)$ is in $C_3^z$. Then $(y,z)$ was not added to $\EB$.

\end{enumerate}

In \ref{case3}, \ref{case6}, and \ref{case7} there is an \ip from $p$ to $q$. Thus Lemma \ref{lemma-path}  tells us there is a path from $p$ to $q$ not longer than $|pa|+\frac{\theta}{\sin\theta}|aq|$.

In \ref{case8}, we have two \ips that meet at $y$. As in the case of a single \ipp, the sum of the lengths of these two paths is not more than $|pa|+\frac{\theta}{\sin\theta}|aq|$. The following lemma proves this claim:

\begin{lemma}\label{lemma-basecase-3}
	
	Consider the edge $(p,r)$ in $\EA$ in the graph $D8(P)$, $r$ in $C_0^p$. Let $(y,z)$ be the last edge in $\C_0^{(p,r)}$, and let  $(y,z)$ be in $C_4^z$. Let $(z,u)$ be an edge in $\EA$ in $C_4^z$. Assume there is an \ip from $p$ to $y$ in $C_0^p$, and an \ip from $z$ to $y$ in $C_4^z$. Let $a$ be the top left corner of $T_{pz}$. We prove an upper bound on the length $\delta(p,z)$ of  $|pa| + \frac{\theta}{\sin{\theta}}|az|.$
	
\end{lemma}

\begin{wrapfigure}[12]{r}{5cm}
	\vspace{-0.5cm}
	\centering
	\includegraphics[page=1, width = 4cm]{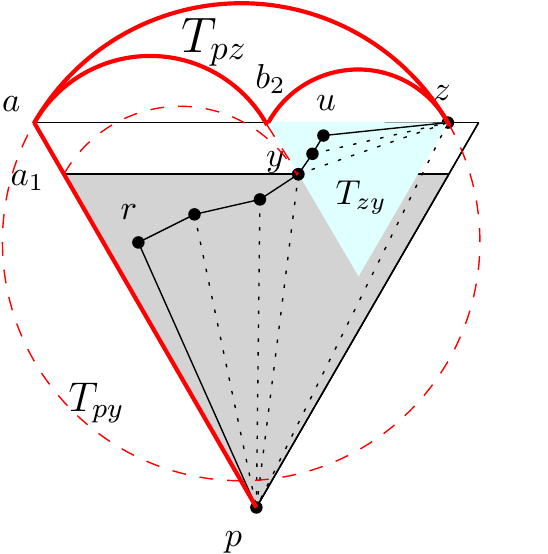}\caption{Concatenating \ipsp.} \label{fig-marked-edge-path}
\end{wrapfigure}

\begin{proof}
	Let $a_1$ be the top left corner of $T_{py}$, and let $b_2$ be the top right corner of $T_{zy}$ (as seen from apex $z$. Note that $T_{zy}$ lies in  $C_z^4$). Since $(y,z)$ is the \emph{last} edge in $\C_0^{(p,r)}$, we note that the ideal path from $p$ to $y$ is to the side of $(p,y)$ that contains $r$ and does not contain $z$. Similarly, the \ip from $z$ to $y$ is to the side of $(y,z)$ that contains $u$ and does not contain $p$. See Figure \ref{fig-marked-edge-path}. By Lemma \ref{lemma-can-path}, the length of the path from $p$ to $z$ in $D8(P)$ is:
	
	\begin{align}
	\delta_{D8(P)}(p,z) &\leq \delta_{D8(P)}(p,y) + \delta_{D8(P)}(z,y) \nonumber\\
	&\leq |pa_1| + \func|a_1s| + |zb_2|+ \func|b_2y| \nonumber\\
	&\leq |pa_1| + \func|a_1s| + |b_2y|+ \func|zb_2| \label{eqn} \\
	&\leq (|pa_1|+|b_2y|)+\func(|a_1y|+|zb_2|) \label{eqn2}\\
	& = |pa| + \func|az| \label{eqn3}
	\end{align}
	
	Inequality \ref{eqn} holds because $\func > 1$, and $|b_2y|\leq |zb_2|$, since $|zb_2|$ is the longest possible line segment in $T_{zy}$.
\end{proof}

In \ref{case5} there is no edge from $y$ to $z$. We prove the length of the path from $p$ to $z$ in \ref{case5} by induction, as part of the main lemma of this section:

\begin{lemma} \label{lemma-inductive-path} 
	Consider the edge $(p,r)$ in $\EA$ in the graph $D8(P)$. Without loss of generality, let $r$ be in $C_0^p$.	Let $a$ and $b$ be the top left corner and top right corner respectively of $T_{pq}$. For any edge $(p,q) \in DT(P)$, there exists a path from $p$ to $q$ in $D8(P)$ that is not longer than $\max\{|pa|+\func|aq|,|pb|+\func|bq|\}$.
\end{lemma}

\begin{proof}
	
	Let $\delta(p,q)$ be the shortest path from $p$ to $q$ in $D8(P)$. We do a proof by induction on the size of the canonical triangle $T_{pq}$.  	
	
	The base case is when $T_{pq}$ is the smallest canonical triangle. One instance of this occurs when there is an \ip from $p$ to $q$, as in \ref{case3}, \ref{case6}, and \ref{case7}. Thus by Lemma \ref{lemma-path}: $$\delta(p,q) \leq \halfspan.$$
	
	The other instance is \ref{case8}, where two \ips meet at a vertex. By Lemma \ref{lemma-basecase-3} we have: $$\delta(p,q) \leq \halfspan.$$
	
	Since $|aq|\leq \max\{|aq|,|bq|\}$, the proof holds in all base cases.
	
	In \ref{case5}, $q$ is the first or last vertex in $\C_0^{(p,r)}$. Since the cases are symmetric, consider when $q$ is the last vertex, and assume it has a neighbour $s$ in $\C_0^{(p,r)}$, such that the canonical edge $(s,q)$ in $N_0^p$ is in $C_0^s$. Thus $(s,q)$ was not added to $\EB$ on a call to $AddCanonical(p,r)$. 
	
	We break down $T_{pq}$ into canonical triangles $T_{ps}$ and $T_{sq}$. Call the upper left corner of $T_{pq}$ $a$, and the upper right corner $b$. Also the upper left corner of $T_{ps}$ is $a_1$, the upper right corner of $T_{sq}$ is $a_2$, the upper right corner of $T_{ps}$ is $b_1$, and the upper right corner of $T_{sq}$ is $b_2$. Since $(s,q)$ is in $C_0^s$, both $T_{ps}$ and $T_{sq}$ must be smaller than $T_{pq}$. 
	
	\begin{figure}%
		\centering
		\subfloat[$|a_1s|\leq|sb_1|, |a_2q|\geq|qb_2|$.] {{		\includegraphics[page=5, width = 4cm]{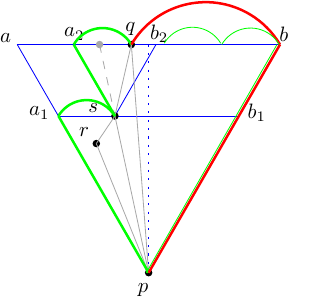} }\label{fig-ind-path}}%
		\qquad
		\subfloat[$|a_1s|\leq|sb_1|, |a_2q|<|qb_2|$.] {{		\includegraphics[page=2, width = 4cm]{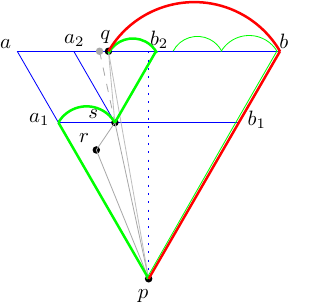} }\label{fig-ind-path2}}%
		\qquad
		\subfloat[$|a_1s|\leq|sb_1|, |qb|<|aq|$.] {{		\includegraphics[page=8, width = 4cm]{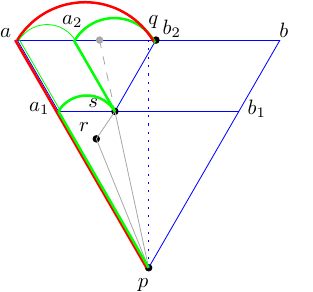} }\label{fig-ind-path3}}%
		\qquad
		\subfloat[$|a_1s|>|sb_1|$.] {{		\includegraphics[page=12, width = 4cm]{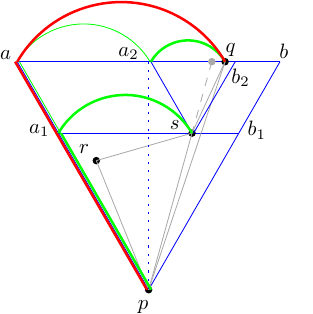} }\label{fig-ind-path4}}%
		\caption{Dark green are the actual paths, light green demonstrates the path is not longer than the red path.}
	\end{figure}
	We note the following facts:
	
	\begin{enumerate}[label=Fact \arabic*:, ref =Fact \arabic*]
		\item \label{fact-duo}$|pa| = |pa_1|+|sa_2|$ and likewise $|pb| = |pb_1|+|sb_2|$
		\item $|ab| = |a_1b_1|+|a_2b_2|$
		\item \label{fact_uno} $|aa_2| = |a_1s|$ and $|b_2b|=|sb_1|$
		\item $q$ is on the line $(a_2,b_2)$
	\end{enumerate}
	
	Without loss of generality, assume the path from $p$ to $s$ is to the side of the line through $p$ and $s$ with $a_1$ (note that we are not assuming that $|a_1s|>|b_1s|$). 
	
	We extend the line $(p,s)$ until it intersects $(a_2,b_2)$ at a point we label $s'$. Since $q$ is the last vertex in $\C_0^{(p,r)}$, $q$ must be to the side of $s'$ closer to $b_2$. 
	
	Since $|pa|=|pb|$ and $|sa_2|=|sb_2|$, it is sufficient to prove:
	\begin{align*}
	|pa_1|+\func|a_1s| + |sa_2| + \func\max\{|a_2q|,|qb_2|\} &\leq |pa| + \func\max\{|aq|,|bq|\}\\
	\end{align*}
	By \ref{fact-duo} this is equivalent to:
	\begin{align*}
	\func|a_1s| + \func\max\{|a_2q|,|qb_2|\}&\leq\func\max\{|aq|,|bq|\}\\
	|a_1s| + \max\{|a_2q|,|qb_2|\} &\leq \max\{|aq|,|qb|\}
	\end{align*}

	We consider two scenarios:
	
	\begin{enumerate}
		\item $|a_1s|\leq |sb_1|$:	There are two sub-cases:
		\begin{enumerate}
			\item $|qb|\geq|aq|$: If $|a_2q|\geq|qb_2|$, then:
			\begin{align*}
			|a_1s|+|a_2q| &\leq |aq| \\
			&\leq |qb|
			\end{align*}
			as required. Otherwise, $|qb_2|>|a_2q|$, thus:
			\begin{align*}
			|a_1s|+|qb_2|&\leq |sb_1|+|qb_2|\\
			&=|qb|
			\end{align*}
			as required.
			\item $|qb|<|aq|$: Together with $|a_1s|\leq |sb_1|$ implies that $|a_2q|>|qb_2|$. See Figure \ref{fig-ind-path4}. Then $|a_1s|+|a_2q| = |aq|$, as required.
		\end{enumerate}
		\item $|a_1s| > |sb_1|$: Since $q$ is radially to the right of $(p,s)$, $|aq|>|qb|$. It is also true that $|a_2q|>|qb_2|$. Thus, using \ref{fact_uno}:
		\begin{align*}
		|a_1s| + |a_2q| & = |aa_2| + |a_2q|\\
		&= |aq|
		\end{align*}
		as required. See Figure \ref{fig-ind-path3}.

	\end{enumerate}
\end{proof}

For an edge $(p,q)$ in $DT(P)$, we have a bound on the length of the path in $D8(P)$. However, this bound is terms of the size of the canonical triangle $T_{pq}$, which is not the same as the Euclidean distance $|pq|$. In the following section we prove that 	$\max\{|pa|+\func|aq|,|pb|+\func|bq|\} \leq \fullspan{|pq|}$.

\subsection{The Spanning Ratio of D8(P)}\label{app:spanning}

\begin{lemma} \label{upper-bound}
	\begin{align*}
	\max\{|pa|+\func|aq|,|pb|+\func|bq|\} \leq \fullspan{|pq|}
	\end{align*}
\end{lemma}

\begin{proof}
	Without loss of generality, we will assume that 
	\begin{align*}
	\max\{|pa|+\func|aq|,|pb|+\func|bq|\}
	=&|pa|+\func|aq|
	\end{align*}
	Let 
	\begin{align*}
	\lambda = \left(\frac{\theta}{\sin\theta}-1\right)(|pq|-|aq|)
	\end{align*}
	
	We will show that:
	
	\begin{align*}
	|pa|+\func|aq| \leq |pa|+\func|aq| + \lambda \leq \fullspan{|pq|}
	\end{align*}
	
	Since $|pq|\geq|pa|$ (by the sine law), and $\frac{\theta}{\sin\theta}>1$, we get $\lambda \geq 0$. Thus

	\begin{align*}
	&|pa|+ \frac{\theta}{\sin\theta}|aq| \\ 
	\leq &|pa|+ \frac{\theta}{\sin\theta}|aq| + \lambda\\
	\end{align*}
	
	It remains to be shown that:
	
	\begin{align*}
	|pa|+ \frac{\theta}{\sin\theta}|aq| + \lambda &\leq \fullspan{|pq|}\\
	|pa|+ \frac{\theta}{\sin\theta}|aq| + \left(\frac{\theta}{\sin\theta}-1\right)(|pq|-|aq|)&\leq \fullspan{|pq|}\\
	|pa|-|pq|+|aq| +\frac{\theta}{\sin\theta}(|aq|+|pq|-|aq|)&\leq \fullspan{|pq|}\\
	|pa|-|pq|+|aq| +\frac{\theta}{\sin\theta}|pq|&\leq |pq| + \frac{\theta}{\sin\theta}|pq|\\
	|pa|-|pq|+|aq| &\leq |pq| \\
	|pa|+|aq| &\leq 2|pq|
	\end{align*}	
	
	Thus we must show that $|pa|+|aq| \leq 2|pq|$ holds true for all values of $\alpha = \angle(apq)$. 
	
	Let $a'$ be the point to the side of $(p,q)$ that contains $a$, such that $\triangle(a'pq)$ is an equilateral triangle. Thus
	$$|pa'|+|a'q|=2|pq|.$$ See Fig. \ref{fig:three}. We will prove that 
	
	$$|pa|+|aq|\leq |pa'|+|a'q|=2|pq|.$$
	\begin{figure}%
		\centering

		\subfloat[]{{\includegraphics[width=4.5cm, page=1 ]{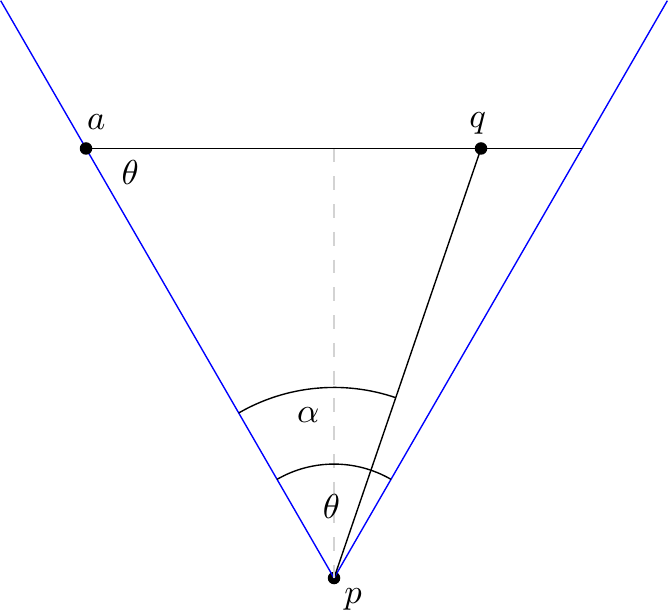} }\label{fig-one}}%
		\qquad
		\subfloat[]{{	\includegraphics[width=4.5cm, page=2 ]{pics/triangle-in.pdf}  }\label{fig:two}}%
		\qquad

		\subfloat[Circle $O_{pqa'}$ also goes through $a$.] {{\includegraphics[width=4.5cm, page=3 ]{pics/triangle-in.pdf}  }\label{fig:three}}%
		\qquad
		\subfloat[The intersection of $O_{pqa'}$ and $E(p,q,d)$.] {{\includegraphics[width=5cm, page=8 ]{pics/triangle-in2.pdf}  }\label{fig:four}}%
		\caption{$|pa|+|aq|\leq |pa'|+|a'q|=2|pq|.$}%
	\end{figure}

	Note that $\angle(paq)=\angle(pa'q)=\theta$. That implies that the circle $O_{pa'q}$ through $p$, $a'$ and $q$ also goes through $a$. See Fig. \ref{fig:three}.

	To better analyze the problem, we rotate, translate, and scale $p$, $q$, $a$ and $a'$ such that $q = (-\sin\theta,0)$, $p= (\sin\theta,0)$, and $a'=(0,-1.5)$. Let $a$ be any point on $O_{pqa'}$ below the line through $p$ and $q$. Let $E(p,q,d)$, where $d=2|pq|$, represent an ellipse with focal points $p$ and $q$ such that for each point $b$ on the boundary of $E(p,q,d)$, $|pb|+|bq|=d = 2|pq|$. Note the center of $O_{pqa'}$ is $(0,-0.5)$, and $O_{pqa'}$ has a radius of $1$. See Fig. \ref{fig:four}. The equation for $O_{pqa'}$ is:
	
	$$x^2 +(y+\frac{1}{2})^2 = 1$$.
	
	The equation for $E(p,q,d)$ is:
	
	\begin{align*}
	\frac{x^2}{a^2}+\frac{y^2}{b^2} &= 1\\
	\frac{x^2}{(2\sin\theta)^2}+\frac{y^2}{\frac{3}{2}^2} &= 1\\
	\frac{x^2}{3}+\frac{4y^2}{9} &= 1
	\end{align*}

	Thus we find the intersection of $O_{pqa'}$ and $E(p,q,d)$ by solving the following system of equations:
	
	\begin{align*}
	x^2 +(y+\frac{1}{2})^2 &= 1\\
	\frac{x^2}{3}+\frac{4y^2}{9}&=1\\
	\end{align*}
	
	This gives us a single solution at $(0,-1.5)$. 
	
	Note that, when $\angle(aqp) = \pi/2$, $|pa|=2$ and $|aq| = 2\cos\theta = 1$. Thus $|pa|+|aq|=3$. We have $2|pq| = 2*(2\sin\theta) \approx 3.46$. Thus when $\angle(aqp) = \pi/2$, $|pa|+|aq|<2|pq|=|pa'|+|a'q|$, which means that $a$ is inside $E(p,q,d)$, which means all of $O_{pqa'}$ is inside $E(p,q,d)$, with the exception of $(0,-1.5)$. Thus for all points $a$ on $O_{pqa'}$, $$|pa|+|aq|\leq |pa'|+|a'q| = 2|pq|.$$ Which implies that:
	
	\begin{align*}
	\delta(p,q)\leq|pa|+\frac{\theta}{\sin\theta}|aq|\leq (1+\frac{\theta}{\sin\theta})|pq|
	\end{align*}
	
	as required.
	\qed \end{proof}

 Using this inequality and Lemma \ref{lemma-inductive-path}, 
the main theorem now follows:

\begin{theorem}\label{theorem:mainspanner8}
	For any edge $(p,q) \in DT(P)$, there is a path in $D8(P)$ from $p$ to $q$ with length at most $\fullspan{|pq|}$, where $\theta = \pi/3$ is the cone width. Thus $D8(P)$ is a $\tspan D_T$-spanner of the complete graph, where $D_T$ is the spanning ratio of the Delaunay triangulation (currently 1.998\cite{xia}).
\end{theorem}

\bibliographystyle{splncs}
\bibliography{mybib}

\begin{thebibliography}{10}

\bibitem{chew}
Chew, P.:
\newblock There is a planar graph almost as good as the complete graph.
\newblock In: Proceedings of the Second Annual Symposium on Computational
  Geometry. SCG '86, New York, NY, USA, ACM (1986)  169--177

\bibitem{dobkin}
Dobkin, D., Friedman, S., Supowit, K.:
\newblock {D}elaunay graphs are almost as good as complete graphs.
\newblock Discrete \& Computational Geometry \textbf{5} (1990)  399--407

\bibitem{keil}
Keil, J., Gutwin, C.:
\newblock Classes of graphs which approximate the complete euclidean graph.
\newblock Discrete \& Computational Geometry \textbf{7} (1992)  13--28

\bibitem{xia}
Xia, G.:
\newblock Improved upper bound on the stretch factor of {D}elaunay
  triangulations.
\newblock In: Proceedings of the Twenty-seventh Annual Symposium on
  Computational Geometry. SoCG '11, New York, NY, USA, ACM (2011)  264--273

\bibitem{bose27}
Bose, P., Gudmundsson, J., Smid, M.:
\newblock Constructing plane spanners of bounded degree and low weight.
\newblock In Möhring, R., Raman, R., eds.: Algorithms - ESA 2002. Volume 2461
  of Lecture Notes in Computer Science.
\newblock Springer Berlin Heidelberg (2002)  234--246

\bibitem{wang23}
Li, X.Y., Wang, Y.:
\newblock Efficient construction of low weight bounded degree planar spanner.
\newblock In Warnow, T., Zhu, B., eds.: Computing and Combinatorics. Volume
  2697 of Lecture Notes in Computer Science.
\newblock Springer Berlin Heidelberg (2003)  374--384

\bibitem{bose17}
Bose, P., Smid, M.H.M., Xu, D.:
\newblock {D}elaunay and diamond triangulations contain spanners of bounded
  degree.
\newblock Int. J. Comput. Geometry Appl. (2009)  119--140

\bibitem{kanj}
Kanj, I.A., Perkovi{\'c}, L., Xia, G.:
\newblock On spanners and lightweight spanners of geometric graphs.
\newblock SIAM Journal on Computing \textbf{39} (2010)  2132--2161

\bibitem{bose-paz}
Bose, P., Carmi, P., Chaitman-Yerushalmi, L.:
\newblock On bounded degree plane strong geometric spanners.
\newblock Journal of Discrete Algorithms \textbf{15} (2012)  16 -- 31

\bibitem{bonichon}
Bonichon, N., Gavoille, C., Hanusse, N., Perkovi{\'c}, L.:
\newblock Plane spanners of maximum degree six.
\newblock In Abramsky, S., Gavoille, C., Kirchner, C., Meyer auf~der Heide, F.,
  Spirakis, P., eds.: Automata, Languages and Programming. Volume 6198 of
  Lecture Notes in Computer Science.
\newblock Springer Berlin Heidelberg (2010)  19--30

\bibitem{degree4}
Bonichon, N., Kanj, I., Perkovi{\'c}, L., Xia, G.:
\newblock There are plane spanners of degree 4 and moderate stretch factor.
\newblock Discrete \& Computational Geometry \textbf{53} (2015)  514--546

\bibitem{euclidean}
Benson, R.:
\newblock {E}uclidean geometry and convexity.
\newblock McGraw-Hill (1966)

\bibitem{bosekeil}
Bose, P., Keil, J.M.:
\newblock On the stretch factor of the constrained {D}elaunay triangulation.
\newblock In: 3rd International Symposium on {V}oronoi Diagrams in Science and
  Engineering, {ISVD} 2006, Banff, Alberta, Canada, July 2-5, 2006, {IEEE}
  Computer Society (2006)  25--31

\end{thebibliography}

\end{document}